\documentclass[12pt]{article}
\usepackage[utf8]{inputenc}
\usepackage[left=1in, right=1in, top=1in]{geometry}
\usepackage[utf8]{inputenc}
\usepackage{appendix}
\usepackage{mathtools}
\usepackage{amsmath}
\usepackage{amssymb}
\usepackage{enumitem}
\usepackage{comment}
\usepackage{amsthm}
\usepackage{setspace}
\usepackage{graphicx}
\usepackage{hyperref}
\usepackage{color}
\usepackage{braket}
\usepackage{bbold}
\usepackage{url}
\usepackage{bibentry}
\usepackage{amsmath}
\usepackage{amsfonts}
\usepackage{clipboard}
\usepackage{cite}
\usepackage{amssymb}
\usepackage{graphicx}
\usepackage{float}
\usepackage{caption}
\usepackage{ulem}
\numberwithin{equation}{section}

\title{FK states in QCD}
\author{ Athira P V, A Manu}
\date{April 2019}

\begin{document}
	
	\baselineskip 24pt
	
	\begin{center}
		
		{\LARGE Classical double copy from Color Kinematics duality : \\ A proof in the soft limit \par}
		
	\end{center}
	
	\vskip .5cm
	\medskip
	
	\vspace*{4.0ex}
	
	\baselineskip=18pt
	
	\centerline{\large \rm  Athira P V$^{a}$, A Manu$^{a}$}
	
	\vspace*{4.0ex}

	\centerline{\it ~$^a$Chennai Mathematical Institute,  SIPCOT IT Park, Siruseri, Chennai, 603103 India} 
	
	\vspace*{1.0ex}

	\vspace*{1.0ex}
	\centerline{\small   \href{mailto:athira@cmi.ac.in}{ \texttt{athira@cmi.ac.in}}, \href{mailto:manu@cmi.ac.in}{\texttt{manu@cmi.ac.in}}, }

	\vspace*{5.0ex}

%
%
%
%
%

\begin{abstract}
Classical double copy is an intriguing relationship between classical solutions to a gravity theory and solutions to classical Yang-Mills equations. Although formally inspired by the double copy relation between (quantum) scattering amplitudes in Yang-Mills and perturbative gravity, a direct proof of the former from the latter continues to be under investigation. In this paper, we attempt to prove classical double copy from the color-kinematics duality symmetry of scalar QCD amplitudes in a restricted setting. That is we consider radiative solutions with classical scattering sources in Yang-Mills theory and perturbative gravity in $D\  \textgreater \ 4$ spacetime dimensions. We show that when the frequency of radiation is much smaller than the characteristic frequency of the process, then at the subleading order in frequency, the classical double copy relating radiative gluon field to radiative gravitational field can be proved from the color-kinematics duality of scalar QCD amplitudes.
\end{abstract}

\newpage
\tableofcontents

\section{Introduction}

Among the many developments in the study of scattering amplitudes in recent years, the discovery of the relationship between scattering amplitudes in gravitational theories with those in gauge theories with color is particularly striking. This relationship manifests itself in two avatars. One goes by the name of  (quantum field theoretic) KLT relations which express amplitudes in perturbative gravity as a ``square" of Yang-Mills amplitudes where the squaring procedure involves convolution of the two amplitudes via a kernel (for a concise derivation and explanation of the KLT formula in quantum field theory, we refer the reader to \cite{cachazo}). The other closely related avatar is known as the double copy which is a consequence of an inherent symmetry present in many theories with color. This symmetry is known as color-kinematics duality and is a symmetry which is not manifest in Feynman diagrammatic expansion of amplitudes but becomes manifest when the amplitudes are written in the so-called BCJ representation \cite{bern}.\\  
On the other hand, there is increasing evidence that in a variety of situations, solutions to classical Yang-Mills equations can be mapped to solutions of gravitational field equations using a ``mapping rule". These rules consist of exchanging the color charges and structure constants of the theory for momenta and three point couplings. This procedure of obtaining classical gravitational solutions from solutions to Yang-Mills field equations is known as the classical double copy\cite{goldberger}. The classical double copy has been used in mapping radiative  gluonic solutions to radiative gravitational field and has also been seen to hold in a variety of other scenarios \cite{ridgway,plefka2,monteiro,white1,white2,nicholson,white3,white4,white5}.\\
A natural question that arises is, can we systematically analyse the double copy and attempt to derive it from first principles ? This may also shed light on domain of it's validity \cite{plefka1}.  In an impressive paper by Shen \cite{shen}, it was shown that one can indeed prove the double copy relations in radiative regime by arranging the classical radiation formulae in such a way that  a ``classical color kinematics duality" became manifest. A proof of the classical double copy starting from scattering amplitudes in QED in 4 space-time dimensions was given recently  in a seminal work \cite{guevara} by  Bautista and Guevara.\\
In spite of these impressive advances, the relationship between scattering amplitudes of gauge theories and gravity poses the following puzzle :  Although the classical double copy appears to be closely related to the quantum double copy for tree level scattering amplitudes, it is still not completely clear what the exact relationship between the two is or to what extent can such a relation between classical solutions in gauge theories and gravity be derived from the color-kinematics duality of quantum amplitudes.\\
In this work we attempt to analyse the origins of the classical double copy from this perspective. The specific classical example we have in mind is the pioneering work by Goldberger and Ridgeway \cite{goldberger}, where it was shown that gluon radiation in a classical scattering process involving color charges can be mapped into gravitational radiation (along with dilaton and axion radiation) using the classical double copy. The question we would like to ask is, to what extent can this double copy be proved from color-kinematics duality of scattering amplitudes. Precisely this question has been asked earlier in \cite{luna} where color-kinematics duality of tree-level amplitudes was used to derive the classical double copy.\\
In this paper we provide a different perspective and proof of the origins of classical double copy from color kinematics duality. Our starting point is the relationship between low frequency classical radiation and quantum soft theorems that was discovered in \cite{ashoke}. It was shown that in $D > 4$ dimensions, the radiative component of low frequency gravitational field in classical gravitational scattering was proportional to the single soft graviton factor. In this work we show that although soft gluon theorems are sufficiently more intricate than soft graviton factors, this result continues to hold in $D > 4$ dimensions. Whence to establish the classical double copy from color-kinematics duality of scattering amplitudes, we need to understand if the graviton and gluon soft factors are related by a double copy derived from the duality.\\
We show that for the case of classical (macroscopic) color charges, which are in a coherent state in the color space, this is indeed true. However our method of proving the classical double copy crucially relies on the soft expansion and hence we call it the ``soft double copy".  In a nut-shell this leads us to an understanding of classical double copy directly from color-kinematics duality atleast for low frequency gluon and graviton radiation. \\
The layout of the paper is as follows : In section \ref{1} we give a brief review of the color-kinematics duality in pure Yang-Mills and matter coupled to Yang-Mills, following it up with the the review on the classical counterpart, the classical double copy in section \ref{2}. In section \ref{GR} we take the soft limit of Goldberger and Ridgeway's results and show that they can be written in terms of the soft factors upto the subleading order in soft momentum. In section \ref{3} we give a brief review of \cite{ashoke}, in which it was shown that low frequency gravitational and electromagnetic radiation can be obtained by (quantum) soft theorems by using saddle point techniques. We repeat the same analysis for low frequency gluon radiation and show that it can be recovered from soft theorems upto the subleading order in soft momentum in section \ref{4}. In section \ref{5} we derive the classical double copy from the color-kinematics duality upto the subleading order in radiation frequency. We then state our results and conclude in section \ref{6}.

\section{Review of Color Kinematics duality} \label{1}

Color-kinematics duality gives a systematic way to relate gauge theory amplitudes to gravitational amplitudes. 
The duality was first discovered in the seminal paper of Bern et al \cite{bern} for tree level amplitudes in Yang-Mills theory (YM). We will give a review of the same and then explain the status of the duality for tree level amplitudes in Quantum Chromodynamics.\\
Consider a $n$-point tree level amplitude of gluons in pure YM. This amplitude can be written in a way that only diagrams with cubic vertices contribute,
\begin{equation}
\label{eq:15}
 \mathcal{A}^{tree}_{n}(1,2,...n) = g^{n-2} \sum_{i=1}^{(2n-5)!!}\frac{c_{i}n_{i}}{D_{i}},
\end{equation}
where 
\begin{itemize}
 \item $c_i$ corresponds to the color factor in the numerator of $i^{th}$ graph and is a product of the structure constants,
 \item $n_i$ corresponds to the kinematical factor in the numerator of $i^{th}$ graph and is a function of the $4$-momenta and polarisation vectors of the particles
 \item and $D_i$ is the propogator of the $i^{th}$ graph.
\end{itemize}
When writing down eq.\eqref{eq:15}, the diagrams with $4$-point vertices have all been absorbed into ones with only cubic interactions. This is done by matching the color structure of the diagrams, multiplying by the required propogators and then redefining the numerators through a gauge transformation.  
In eq.\eqref{eq:15}, from the full set of diagrams one can find sets of three diagrams whose numerators ($n_{i}$) satisfy the same Jacobi identity as the corresponding color factors ($c_{i}$),
\begin{align} 
  & c_{i}\pm c_{j}\pm c_{k}=0 \ \ , \    n_{i}\pm n_{j}\pm n_{k}= 0,  \label{eq:16}\\             
  & c_{i} \rightarrow -c_{i} \ \ , \    n_{i} \rightarrow -n_{i}.       \label{eq:17}
 \end{align} 
The second relation tells us that flipping of two legs leads to a minus sign in both the color factors and the kinematical numerator factors.\\
These relations are called the color-kinematics duality.\\ When every numerator in an $n$-point tree level amplitude of gluons satisfies at least one color-kinematics relation we will say that one has found a BCJ representation of the amplitude. Once a BCJ representation has been found we do the (quantum) double copy, which is to substitute the color factors with the respective numerator factors ($ c_{i} \rightarrow n_{i} $) in 
\eqref{eq:15}. The double copy, then gives us the $n$-point tree level graviton amplitude, 
\begin{equation}
\label{eq:19}
  \mathcal{M}^{tree}_{n}(1,2...n) = \bigg(\frac{\kappa}{2}\bigg)^{n-2}\sum_{i=1}^{(2n-5)!!}\frac{n_{i}n_{i}}{D_{i}}
\end{equation}
This can be verified to be the correct $n$-point tree level amplitude for gravitons by Feynman diagrams\footnote{To get a gravitational amplitude one has to do the substitutions for the coupling constants also, which is  
\begin{equation}
 g\rightarrow \frac{\kappa}{2}.
\end{equation}
}.
It has further been proven in \cite{bern1}, using the BCFW recursion relations\cite{bcf,witten}, that color-kinematics duality implies the double copy property.\\ 
The status of the color-kinematics duality with inclusion of massive matter in arbitrary representation and arbitrary spin has seen significant progress in recent times. In the seminal work \cite{ochirov}, authors proved the 
color kinematics duality for tree level amplitudes in QCD.  These results were used in many future works. In ref.\cite{johanssonochirov},it was shown that by coupling fermions (in the fundamental representation) to gluons and treating them as ghost fields one could construct pure gravity and supergravity amplitudes in $D = 4$, via the color-kinematics duality.\\
When the matter field is colored scalar, color-kinematics duality was proved in \cite{vera,johansson}. This will be the case of interest for the current paper and we will review the derivation of this duality in detail in \eqref{5}. 
Recently, a completely general characterization of color-kinematics duality and the accompanying double copy has been established for generic matter content coupling to gluons\cite{ochirov2}.

\section{Review of the classical double copy} \label{2}

In recent years, the relationship between gauge theories and gravitational theories has been studied extensively at the level of classical solutions using different proposals for the classical  double copy. For instance it was shown in \cite{monteiro} that certain classes of spacetimes called Kerr-Schild geometries can be obtained from abelian solutions of Yang-Mills theory by substituting the color charge in the YM solutions with the null vector present in the Kerr-Schild coordinates. Then in \cite{goldberger}, the classical double copy was generalised to map radiative solutions in gauge theories to the ones in gravitational theories.\\
Specifically the authors in \cite{goldberger} proposed a set of substitutions that can be used to map the classical radiative gauge field, generated by the scattering of $N$ massive spinless colored particles, to the far field metric generated by the scattering of $N$ massive spinless gravitating bodies at leading order in the coupling constants.\\
In the following sections we will review the classical double copy \cite{goldberger}. 

\subsection{Classical scattering of colored particles}
The authors considered the scattering of massive colored particles with positions $x^{\mu}_{i}(\tau)$ and color charges $c^{a}_{i}(\tau)$ ($i=1,2,..N$) interacting through the Yang-Mills field $A^{a}_{\mu}(x)$, in the worldline formalism. The particles are in the adjoint representation of the Lie algebra of the gauge group $SU(N)$. The equations of motion are
\begin{align}
 \Box A_{a}^{\mu}(x) = g\tilde{J}^{\mu}_{a}(x), \\
 \frac{dc^{a}_{i}(\tau_{i})}{d\tau_{i}} = g \sum_{j\neq i}^{N} f^{abc}A^{b}_{j\mu}(x(\tau_{i}))v^{\mu}_{i}(\tau_{i})c^{c}_{i}(\tau_{i}), \label{eq:37} \\
 \frac{d^{2}x^{\mu}(\tau_{i})}{{d\tau}^{2}_{i}} = g\sum_{j\neq i}^{N} c^{a}_{i}(\tau)F^{\mu\nu a}_{j}(x(\tau_{i}))v_{\nu i}(\tau_{i}) \label{eq:38}.
\end{align}
The conserved current $\tilde{J}^{a}_{\mu}(x)$ consists of two parts. The first is from the matter and second from the gauge field mediating the interaction between the particles,
\begin{align}\label{eq:103}
 \tilde{J}^{\mu a} = J^{\mu a} + f^{abc}A^{b}_{\nu}(\partial^{\nu}A^{\mu c} - F^{\mu\nu}_{c}),
\end{align}
 with the matter current given by
\begin{equation} \label{eq:42}
 J^{\mu a}(x) = \sum_{i=1}^{N} \int d\tau_{i} c^{a}_{i}(\tau_{i})v^{\mu}_{i}(\tau_{i})\delta^{D}(x-r_{i}(\tau_{i})).
\end{equation}
The expression for the radiative gauge field in arbitrary space time dimensions is
\begin{equation} \label{eq:40}
 \lim_{r\rightarrow \infty}A^{\mu}_{a}(x) = \frac{g}{4\pi r^{(D-2)/2}}\int \frac{d\omega}{2\pi} e^{-i\omega t} \tilde{J}^{\mu}_{a}(k).
\end{equation}
Here $k = \omega\hat{k} = \omega(1,\hat{n})$. The $\hat{n} = \vec{x}/r$ is the direction in space where the radiation is measured.\\
The idea is to solve the equations of motion using perturbation theory. This was done by ensuring that the deflection from the initial trajectories of the particles, compared to the impact parameter, is small. \\ 
At the zeroth order, $\mathcal{O}(g^{0})$, the particles move at uniform velocity and constant color charges,
\begin{align}
 x^{\mu}_{i}(\tau_{i}) \approx b^{\mu}_{i} + v_{i}^{\mu}\tau_{i},\\
 c^{a}_{i}(\tau) \approx c^{a}_{i}.
\end{align}
The $v^{\mu}_{i}$, $c^{a}_{i}$ and $b^{\mu}_{i}$ are the initial velocity, color charge and the position in space associated with the $i^{th}$ particle. The gauge field produced by the particles at this order, $\mathcal{O}(g)$, is a Coulombic gauge field.  Hence, at this order there is no radiation.\\
Denoting the change in color charge as $\bar{c}^{a}(\tau)$ and the deflection in the path of the initial trajectory as $z^{\mu}(\tau)$, the color charge and the trajectory of the particle at arbitrary proper time is then given by
\begin{align}
 x^{\mu}_{i}(\tau) = b^{\mu}_{i} + v^{\mu}_{i}\tau + z^{\mu}_{i}(\tau),\\
 c^{a}_{i}(\tau) = c^{a}_{i} + \bar{c}^{a}_{i}(\tau).
\end{align}
The Coulombic gauge field can be used to obtain the the change in color charge and deflection in the path of the particles using eq.\eqref{eq:37} and eq.\eqref{eq:38}.
The differential equations satisfied by these quantities are
\begin{align}
 \frac{d\bar{c}_{i}^{a}(\tau)}{d\tau} = -g^{2}\sum_{j \neq i} (v_{i}\cdot v_{j})f^{abc}c^{b}_{j}c^{c}_{i}\int_{l} (2\pi) \delta(l\cdot v_{j})\frac{e^{il\cdot (b_{ij}+v_{i}\tau)}}{l^{2}}, \label{eq:200}\\
 \frac{d^{2}z _{i}^{\mu}(\tau)}{d\tau} = -ig^{2}\sum_{j \neq i} (c_{i}\cdot c_{j})\int_{l} (2\pi) \delta(l\cdot v_{j})\frac{e^{il\cdot (b_{ij}+v_{i}\tau)}}{l^{2}}\big[(v_{i}\cdot v_{j})l^{\mu}-(v_{i}\cdot l)v^{\mu}_{j}\big]. \label{eq:201}
\end{align}
Here $b^{\mu}_{ij} = b^{\mu}_{i} - b^{\mu}_{j}$, is the impact parameter.
Using these expressions one then obtains the conserved current in Fourier space.
The expression for $\tilde{J}^{\mu}_{a}(k)$ at $\mathcal{O}(g^{2})$ is 
\begin{multline} \label{eq:45}
 \tilde{J}^{\mu a}(k) = g^{2} \sum_{\substack{i,j \\ i \neq j}} \int_{l_{i},l_{j}} \mu_{i,j}(k) \bigg[ \frac{c_{i}\cdot c_{j}}{m_{i}}\frac{l_{i}^{2}}{k\cdot v_{i}}c_{i}^{a} \bigg\{-v_{i}\cdot v_{j}\bigg( l^{\mu}_{j} - \frac{k\cdot l_{j}}{k\cdot v_{i}} v_{i}^{\mu}\bigg) + (k\cdot v_{i}) v_{j}^{\mu} - (k\cdot v_{j}) v^{\mu}_{i}\bigg\}\\
 +if^{abc}c^{b}_{i}c^{c}_{j}\bigg\{ 2(k\cdot v_{j})v^{\mu}_{i} - (v_{i}\cdot v_{j})l_{i}^{\mu} + (v_{i}\cdot v_{j})\frac{l^{2}_{i}}{k\cdot v_{i}}v^{\mu}_{i}\bigg\} \bigg].
\end{multline}
The integrals come from the static gauge field mediating the interaction of the particles.
The functional form of $\mu_{i,j}(k)$ is
\begin{equation} \label{mudef}
 \mu_{i,j}(k) = \bigg[(2\pi)\delta(v_{i}\cdot l_{i})\frac{e^{il_{i}\cdot b_{i}}}{l^{2}_{i}}\bigg]\bigg[(2\pi)\delta(v_{j}\cdot l_{j})\frac{e^{il_{j}\cdot b_{j}}}{l^{2}_{j}}\bigg](2\pi)^{D}\delta^{D}(k-l_{i}-l_{j}).
\end{equation}
Using the conserved current, eq.\eqref{eq:45}, one can obtain radiative gauge field at $\mathcal{O}(g^{3})$ from eq.\eqref{eq:40}.

\subsection{Classical gravitational scattering}

A similar analysis was then done to calculate the far field metric generated by the scattering of massive spinless classical particles whose interactions are mediated by gravitons and dilatons (scalar field). The equation of motion for a point particle in dilaton gravity is
\begin{equation} \label{dgraveoms}
 \frac{dp^{\mu}(\tau)}{d\tau} = - [\Gamma^{\mu}_{\rho\nu}v^{\rho}(\tau) + (v^{\mu}\partial_{\nu} - v_{\nu}\partial^{\mu})\phi(x(\tau))]p^{\nu}(\tau).\\ 
\end{equation}
The $\phi$ here is the dilaton. The metric is then perturbatively expanded around the flat background, $g_{\mu\nu} = \eta_{\mu\nu} + h_{\mu\nu}$. Then by choosing the deDonder gauge, the solution to the equation of motion for gravity can be written down. From there one calculates the far field radiation field which is given by  
\begin{equation}
 \lim_{r\rightarrow \infty} h_{\mu\nu}(t,\vec{n}) = \frac{4G_{N}}{r^{(D-2)/2}} \int \frac{d\omega}{2\pi} e^{-i\omega t} \tilde{T}_{\mu\nu}(k).
\end{equation}
The stress energy tensor for the matter part is given by
\begin{equation} \label{eq:41}
 T^{\mu\nu}(x) = \sum_{i=1}^{N}\int d\tau p^{\mu}_{i}(\tau_{i})v^{\nu}_{i}(\tau_{i})\delta^{D}(x-r_{i}(\tau_{i})).
\end{equation}
One then solves the equations of motion in eq.\eqref{dgraveoms} following the same steps as for the gauge theory setup. Finally one obtains the conserved stress energy tensor, $\tilde{T}^{\mu\nu}(k)$ at the order $\mathcal{O}(M_{pl}^{-(D-2)})$ which is 
\begin{multline} \label{eq:101}
 \tilde{T}^{\mu\nu}(k) = \frac{1}{2M_{pl}^{D-2}}\sum_{\substack{i,j \\ i \neq j}} m_{i}m_{j}\int_{l_{i},l_{j}} \mu_{i,j}(k)\bigg[ (v_{i}\cdot v_{j})^{2}l_{i}^{\mu}l_{j}^{\nu} + (v_{i}\cdot v_{j})\eta^{\mu\nu} \bigg\{ \frac{1}{2}(v_{i}\cdot v_{j})l^{2}_{i} + (k\cdot v_{i})(k\cdot v_{j})\bigg\}\\
 - 2(v_{i}\cdot v_{j})\bigg((v_{i}\cdot v_{j})\frac{l^{2}_{i}}{k\cdot v_{i}} + 2(k\cdot v_{j})\bigg)l_{i}^{\mu}v^{\nu}_{i} - 2\big((k\cdot v_{i})(k\cdot v_{j}) + (v_{i}\cdot v_{j})l^{2}_{i}\big)v_{i}^{\mu}v^{\nu}_{j}\\
 + \bigg\{ (v_{i}\cdot v_{j})\frac{l^{2}_{i}}{(k\cdot v_{i})}\big((v_{i}\cdot v_{j})(k\cdot l_{i}) + 2(k\cdot v_{i})(k\cdot v_{j})\big) + 2(k\cdot v_{j})^{2}\bigg\}v_{i}^{\mu}v_{j}^{\nu}\bigg].
\end{multline}
The $M^{D-2}_{pl}$ is the Planck mass in $D$ spacetime dimensions. The eq.\eqref{eq:101} can then be used to obtain the far field metric by using eq.\eqref{dgraveoms}. This stress tensor contains contributions from both the dilaton and the graviton.

\subsection{Classical double copy}

It was shown in \cite{goldberger} that the two classical solutions namely the conserved current at $\mathcal{O}(g^{2})$ on the gauge theory and the conserved stress energy tensor at $\mathcal{O}\big(\frac{1}{M_{pl}^{D-2}}\big)$ can be mapped from the gauge theory to the gravity side by the following set of substitutions,
\begin{align}
 c_{a} \rightarrow p_{\mu} \label{eq:43},\\
 if^{abc} \rightarrow \Gamma^{\mu\nu\rho} = -\frac{1}{2}\big[\eta^{\mu\nu}(p-q)^{\rho} + 
 \eta^{\nu\rho}(q-r)^{\mu} + \eta^{\rho\mu}(r-p)^{\nu}\big], \\
 g \rightarrow \frac{1}{2M_{pl}^{D/2-1}}.  
 \end{align}
The first substitution states that the classical color charge of the particle on the gauge theory side has to be replaced with the $4$-momentum of the same particle. The motivation for this can be inferred by seeing the similarities in the evolution equation for the color charge and the geodesic equation.
Note that the matter current in eq.\eqref{eq:42} can be mapped to the stress tensor in eq.\eqref{eq:41} for the matter particles using this substitution.\\
The second substitution is slightly less obvious. It substitutes structure constant $f^{abc}$ with the $3$-point vertex for Yang-Mills theory (as opposed to gravity!). One possible motivation for this rule comes from the fact that it replaces the full $3$- point vertex of YM theory, namely $f^{abc}\ \Gamma^{\mu\nu\rho}$ with $3$- point vertex of perturbative gravity.\\ 
In a series of subsequent papers \cite{goldberger1,ridgway,li,prabhu}, more evidence was produced in support of the classical double copy. 

\section{Soft limit of gluon and gravitational radiation} \label{GR}

In this section we compute the soft limit of the (Fourier transform of) radiative gauge field and gravitational field given in eq.(\ref{eq:45}) and eq.(\ref{eq:101}) respectively. An efficient approach to compute soft radiation from scattering amplitudes in $D\  \textgreater \ 4$ dimensions, was outlined in \cite{ashoke}. It can be easily verified that for gluon radiation, this approach leads to the result precisely analogous to the electromagnetic soft factor derived in \cite{ashoke} with the only difference being that the electric charges of scattering particles are replaced by the color charges which evolve along the trajectory of the particles. \\
However as our goal is to derive the classical double copy relations relating $\tilde{J}_{\mu}^{a}$ and $\tilde{T}_{\mu\nu}$ given in eq.(\ref{eq:45}),eq.(\ref{eq:101}), we consider the soft limit of their expressions for the radiative fields. As we show below, this leads to the expected result. Namely that low frequency classical radiation is given by sum of leading and subleading classical soft factors. We will first consider the gluon radiation and then discuss the results for the gravitational field. \\
Expanding eq.\eqref{eq:45} upto the leading order in soft momentum, we get
\begin{multline} \label{eq:46}
 \tilde{J}^{\mu}_{a}(k) = g^{2} \sum_{\substack{i,j \\ i \neq j}} \int_{l_i,l_j} \mu_{i,j}(k=0) \bigg[ \frac{c_{i}\cdot c_{j}}{m_{i}}\frac{l_{i}^{2}}{k\cdot v_{i}}c_{i}^{a} \bigg\{-v_{i}\cdot v_{j}\bigg( l_{j} - \frac{k \cdot l_{j}}{k\cdot v_{i}} v_{i}^{\mu}\bigg)\bigg\}\\
 +if^{abc}c^{b}_{i}c^{c}_{j}(v_{i}\cdot v_{j})\frac{l^{2}_{i}}{k\cdot v_{i}}v^{\mu}_{i} + \mathcal{O}(1) \bigg].
\end{multline}
To write this expression in terms of the leading soft gluon factor, we need two expressions. The first is the total change in color charge of the scattering particles and the total deviation from their initial straight line trajectories. Integrating the equations \eqref{eq:200} and \eqref{eq:201} over the whole trajectory of the particle, we obtain  
\begin{align} \label{eq:52}
 \bar{c}^{a}_{i}(\tau_{+}) = g^{2} \sum_{j \neq i} f^{abc} c^{b}_{i}c^{c}_{j} \int_{l_{i},l_{j}} \mu_{i,j}(k=0) \ l^{2}_{i}, \\
 \frac{dz^{\mu}_{i}(\tau_{+})}{d\tau} = ig^{2} \sum_{j \neq i}  (v_{i}\cdot v_{j})\frac{c_{i}\cdot c_{j}}{m_{i}} \int_{l_i,l_j} \mu_{i,j}(k=0)\ l^{2}_{i}\ l^{\mu}_{j}. \label{eq:115}
\end{align}
The left hand side denotes the changes in the color charge ($\bar{c}^{a}_{i}(\tau_{+})$) and velocity ($\frac{dz^{\mu}_{i}(\tau_{+})}{d\tau}$) evaluated as $\tau \rightarrow \infty\ (\tau_{+})$. Using these two expressions it is not difficult to see that eq.\eqref{eq:46} can be written as follows,
\begin{equation} \label{eq:47}
 \tilde{J}^{\mu}_{a}(k) = i\sum_{i=1}^{N} \bigg[ \frac{c^{a}_{i}}{k\cdot v_{i}} \frac{dz_{i}^{\mu}(\tau_{+})}{d\tau} -	 \frac{c^{a}_{i}v^{\mu}_{i}}{(k\cdot v_{i})^{2}} \frac{k\cdot dz_{i}(\tau_{+})}{d\tau} + \bar{c}^{a}(\tau_{+})\frac{v^{\mu}_{i}}{k\cdot v_{i}} + \mathcal{O}(1) \bigg].
\end{equation}
We denote the final color charge and the final $4$-velocity of a particle as
\begin{align}
 c^{+}_{a} = c^{-}_{a} + \bar{c}_{a}(\tau_{+})\ \ \ \ \ \ \ \ \ ,~c^{-}_{a} = c_{a},\\
 v^{+}_{\mu} = v^{-}_{\mu} + \frac{dz_{\mu}(\tau_{+})}{d\tau} \ \ \ \ \ \ \ \ \ , ~v^{-}_{\mu} = v_{\mu}.
\end{align}
Using these definitions and the fact that the net deflection is small,\footnote{This just means that the change in the particle's velocity,\eqref{eq:115}, is negligible compared to its initial velocity,
\begin{equation}
 \frac{k\cdot v^{+} - k\cdot v^{-}}{k\cdot v^{-}} = \frac{k\cdot dz/d\tau}{k\cdot v^{-}} << 1 
\end{equation}} eq.\eqref{eq:47} can be rewritten only in terms of the final and initial $4$-velocities and color charges of the particles. Hence,
\begin{equation}
 \tilde{J}^{\mu a}(k) = i\sum_{i=1}^{N} \bigg[ c^{+}_{ai}\frac{v^{+\mu}_{i}}{k\cdot v^{+}_{i}} - c^{-}_{ai}\frac{v^{-\mu}_{i}}{k\cdot v^{-}_{i}} + \mathcal{O}(1) \bigg].
\end{equation}
Hence the radiative gauge field for the above conserved current can be computed and is 
\begin{align} \label{eq:120}
 A^{\mu}_{a}(\omega,\vec{x}) = & \frac{ig}{r^{(D-2)/2}} e^{i\omega r} \sum_{i=1}^{N} \bigg[ c^{+}_{ai}\frac{v^{+\mu}_{i}}{k\cdot v^{+}_{i}} - c^{-}_{ai}\frac{v^{-\mu}_{i}}{k\cdot v^{-}_{i}} + \mathcal{O}(1) \bigg]\\
  = & \frac{ig}{r^{(D-2)/2}} e^{i\omega r} S^{(0)\mu}_{a}(\epsilon,k),
\end{align}
where  
\begin{equation}
 S^{(0)\mu}_{a}(\epsilon,k) = \sum_{i=1}^{N} \bigg[ c^{+}_{ai}\frac{v^{+\mu}_{i}}{k\cdot v^{+}_{i}} - c^{-}_{ai}\frac{v^{-\mu}_{i}}{k\cdot v^{-}_{i}}\bigg]
\end{equation}
is the leading soft gluon factor.\\
Low frequency limit of the radiative gravitational field follows the analogous analysis.  We start with, 
\begin{equation} \label{eq:102}
\tilde{T}^{\mu\nu}(k) = \frac{G_{N}}{4} \sum_{\substack{i,j \\ i\neq j}} m_{i}m_{j} (v_{i}\cdot v_{j})^{2}\int_{l_{i},l_{j}}\frac{l^{2}_{i}}{k\cdot v_{i}} \big[v^{\mu}_{i}l^{\nu}_{j} + v^{\nu}_{i}l^{\mu}_{j} - \frac{k\cdot l_{j}}{k\cdot v_{i}}v_{i}^{\mu}v_{i}^{\nu} + \mathcal{O}(1)\big].
\end{equation}
We first consider the leading soft limit of $\tilde{T}^{\mu\nu}(k)$.
Using
\begin{equation}
 \frac{dz^{\mu}_{i}(\tau_{+})}{d\tau} = \frac{-iG_{N}}{4} \sum_{j\neq i} m_{j}(v_{i}\cdot v_{j})^{2}\int_{l_i,l_j} \mu_{i,j}(k=0)\ l^{2}_{i}\ l^{\mu}_{j},
\end{equation}
eq.\eqref{eq:102} in the small deflection approximation can be written as
\begin{equation}
 \tilde{T}^{\mu\nu}(k) = i \sum_{i=1}^{N} m_{i} \bigg[ \frac{v_{i}^{\mu}}{k\cdot v_{i}}\frac{dz^{\nu}_{i}(\tau_{+})}{d\tau} + \frac{v_{i}^{\nu}}{k\cdot v_{i}}\frac{dz^{\mu}_{i}(\tau_{+})}{d\tau} - \frac{v_{i}^{\mu}v_{i}^{\nu}}{(k\cdot v_{i})^{2}}\frac{k\cdot dz_{i}(\tau_{+})}{d\tau} + \mathcal{O}(1)\bigg].
\end{equation}
Denoting the final $4$-velocity of a particle as
\begin{equation}
 v^{+\mu} = v^{-\mu} + \frac{dz^{\mu}(\tau_{+})}{d\tau} \ \ \ \ \ \ \ \ \ ,~v^{- \mu }_{i} = v^{\mu}_{i},
\end{equation}
the above equation can be written as 
\begin{equation}
 \tilde{T}^{\mu\nu}(k) = i \sum_{i=1}^{N} m_{i} \bigg[ \frac{v^{+\mu}_{i}v^{+\nu}_{i}}{k\cdot v^{+}_{i}} - \frac{v^{-\mu}_{i}v^{-\nu}_{i}}{k\cdot v^{-}_{i}} + \mathcal{O}(1) \bigg].
\end{equation}
The corresponding far field metric is then given by
\begin{align} \label{eq:121}
 h^{\mu\nu}(\omega,\vec{x}) = & \frac{i}{r^{(D-2)/2}} e^{i\omega r} \sum_{i=1}^{N} m_{i} \bigg[ \frac{v^{+\mu}_{i}v^{+\nu}_{i}}{k\cdot v^{+}_{i}} - \frac{v^{-\mu}_{i}v^{-\nu}_{i}}{k\cdot v^{-}_{i}} + \mathcal{O}(1) \bigg],\\
  = & \frac{i}{r^{(D-2)/2}} e^{i\omega r} S^{(0)\mu\nu}_{GR}(\epsilon,k).
\end{align}
The 
\begin{equation}
 S^{(0)\mu\nu}_{GR}(\epsilon,k) = \sum_{i=1}^{N} \bigg[ \frac{p^{+\mu}_{i}v^{+\nu}_{i}}{k\cdot v^{+}_{i}} - \frac{p^{-\mu}_{i}v^{-\nu}_{i}}{k\cdot v^{-}_{i}}\bigg],
\end{equation}
is the single leading soft graviton factor.\\
Comparing eq.\eqref{eq:120} and eq.\eqref{eq:121}, it is easy to see that the two expressions are related to each other via the first of the classical double copy substitutions,\eqref{eq:43}.\\
One can extend the above computation for radiation field to subleading order in radiation frequency. 
As we show in the appendix (\ref{slglu}), we can extend the analysis to the subleading order as
\begin{equation} \label{subgauge}
A^{\mu}_{a}(\omega,\vec{x})\ =\ \frac{ig}{r^{(D-2)/2}} e^{i\omega r} \ \big[S^{(0)\mu}_{a}(\epsilon,k)\ +\ S^{(1)\mu}_{a}(\epsilon, k)\big].
\end{equation}
Where the  classical soft gluon factor at the subleading order is given by,
\begin{equation} \label{eq:subleading}
S^{(1)\mu}_{a}(\epsilon, k)\ =\ \sum_{i=1}^{N} \bigg[ c^{+}_{ai}\frac{\bold{J}^{+\mu\nu}_{i}\ k_{\nu}}{k\cdot v^{+}_{i}} - c^{-}_{ai}\frac{\bold{J}^{-\mu\nu}_{i}\ k_{\nu}}{k\cdot v^{-}_{i}} \bigg].\
\end{equation}
$\bold{J}^{\pm}_{i}$ are the initial and final orbital angular momenta of the $i^{th}$ particle.\\ 
As the computation is rather analogous to the one given in appendix (\ref{slglu}) for the gluon radiation, we simply give the final result which matches with the one given in \cite{ashoke}
\begin{equation} \label{subgrav}
h^{\mu\nu}(\omega,\vec{x}) = \mathcal{N}e^{i\omega r} \sum_{i=1}^{N} \bigg[ \frac{p^{+\mu}_{(i)}p^{+\nu}_{(i)}}{p^{+}_{(i)}\cdot k} - \frac{p^{-\mu}_{(i)}p^{-\nu}_{(i)}}{p^{-}_{(i)}\cdot k} + i\bigg( \frac{p^{+\mu}_{(i)}k_{\gamma}\bold{J}^{+\gamma\nu}_{(i)}}{p^{+}_{(i)}\cdot k}  - \frac{p^{-\mu}_{(i)}k_{\gamma}\bold{J}^{-\gamma\nu}_{(i)}}{p^{-}_{(i)}\cdot k}\bigg) \bigg].
\end{equation}
\\
In section \ref{Extremisation} we show that the low frequency radiative gauge field given in \eqref{eq:120} can also be computed from the soft gluon theorem in scalar QCD.  This will set the stage for us to prove the classical double copy relations at subleading order in the frequency from color-kinematics duality for scalar QCD amplitudes. 

\section{Gravitational radiation from soft theorems} \label{3}

In this section we review the results of \cite{ashoke} in which the authors considered classical gravitational scattering involving two massive objects in the initial state. As was shown, for a classical scattering in which either ({\bf a}) the so called probe-scatterer approximation is valid\footnote{We say that in a given $2\rightarrow\ 2 + N$, the probe-scatterer limit is valid if mass of one of the initial objects (called probe) is much smaller then mass of the other object called scatterer.} or ({\bf b}) the impact parameter ($b$) is much larger then the Schwarzschild radii of the scattering objects, low frequency classical radiation could be computed by the (quantum) multiple soft graviton theorem. 

In the large impact parameter approximation, their analysis showed that upto sub-subleading order in soft expansion, the radiative component of $h_{\mu\nu}$ is proportional to sum of classical soft factors at leading, subleading and sub-subleading order. 

More in detail, using the multiple soft graviton theorem, it was proved in \cite{ashoke} that the soft expansion of  radiative component of the gravitational field was given by

\begin{equation}\label{gclso}
h_{\mu\nu}(\omega,\vec{x}) = \mathcal{N}e^{i\omega R} S^{\textrm{GR}}_{\mu\nu}(\epsilon,k),
\end{equation}
where $\omega$ is the low frequency ($\omega\ <<\ b^{-1}$), $\vert\vec{x}\vert\ =\ R$ and $\mathcal{N} =\frac{1}{2\omega} \bigg(\frac{\omega}{2\pi R}\bigg)^{(D-2)/2} $. The R.H.S. is given by  
\begin{equation}
\begin{array}{lll}
S^{\textrm{GR}}_{\mu\nu}(\epsilon,k) = \sum_{i=1}^{N} \bigg[ \frac{p_{(i)\mu}p_{(i)\nu}}{p_{(i)}\cdot k} + i\frac{p_{(i)\mu}k^{\gamma}\bold{J}_{(i)\gamma\nu}}{p_{(i)}\cdot k}\ - \frac{1}{2}\frac{1}{p_{(i)}\cdot k}k^{\gamma}k^{\rho}\bold{J}_{(i)\mu\gamma}\bold{J}_{(i)\nu\rho}\bigg].
\end{array}
\end{equation}
Here the $x^{\mu}$s denote the asymptotic trajectories of the classical object and $\bold{J}_{(i)\mu\nu}$ is the total angular momentum of the $i^{th}$ particle involved in the scattering. 

This derivation of soft gravitational radiation  in \cite{ashoke} relied on the fact that although the multiple soft graviton theorem has a rather intricate structure in quantum gravity, if the impact parameter of scattering is large then the theorem simplifies drastically.  The simple form of multiple soft graviton theorem followed by a saddle point analysis leads to classical formulae for radiative field from quantum soft factors. 

Large $b$ approximation is precisely the scenario in which classical gluon as well as gravitational scattering in \cite{goldberger} was analysed and indeed, as we reviewed above in eqns. \eqref{subgauge},\eqref{subgrav}, soft $\omega$ expansion of results of \cite{goldberger} matched with those of \cite{ashoke}. 

Whence we would now like to derive low frequency gluon radiation using multiple soft gluon theorem. In contrast to the soft graviton theorems, the soft gluon factorization is only valid till subleading order and hence  we will restrict our analysis to $O(\omega^{0})$. We will briefly comment on the sub-subleading terms of the gluonic radiative field in the conclusion section. 

\section{Low frequency gluon radiation from soft theorems} \label{4}

In this section we will derive the low frequency gluon radiation from (quantum) soft gluon theorem.\\
The basic idea is to consider the scattering amplitude involving two incoming colored particles which scatter into two outgoing colored particles and N soft gluons. We work in large impact parameter approximation.  
Following the derivation in \cite{ashoke}, we first find the probability for the emission of the N soft gluons in a given bin 
and take the classical limit by finding the number of emitted soft gluons in a given bin for which the probability distribution peaks.
For the gluons this is easier said than done even at leading order in soft frequency. This is due to two complications. 
\begin{itemize}
\item As gluons are in the adjoint of the gauge group, the soft factor is in fact an operator in the color space. This is simply because when a gluon is emitted from an external state like a quark, the color of the final state is different from the color of the initial state. This is independent of the frequency of the gluon and hence such a color rotation survives even in the soft limit. Of course this feature is extensively analysed in the \textcolor{blue}{QCD} literature \cite{berends}. 
\item Even at leading order the multiple soft gluon theorem has the so called contact terms which further complicates the classical (saddle point) analysis. We note that both these features are reminiscent of the subleading multiple soft graviton theorem where soft factor is an operator in momentum space (due to the presence of angular momenta) and has contact terms with a universal but rather intricate structure.

\end{itemize}
As we will see below, the nature of contact terms is such that for the classical question of interest they vanish and the analysis simplifies considerably.

\subsection{Multiple soft gluon expansion}

We thus consider the $(4+N)$-point amplitude of emission of $N$ soft gluons from a scattering of two massive scalar particles. The two scalar particles have different masses and live in the adjoint representation of the Lie algebra of the group $SU(N)$. The process can be written as 
\begin{multline} \label{eq:50}
 \Phi(M,p_{1},a_{1}) + \phi(m,p_{2},a_{2}) \rightarrow \Phi(M,p_{3},a_{3}) + \phi(m,p_{4},a_{4}) \\ 
 + g_{1}(\epsilon_{1},k_{1},b_{1}) + g_{2}(\epsilon_{2},k_{2},b_{2})+...+ g_{N}(\epsilon_{N},k_{N},b_{N}).
\end{multline}
The $\Phi$ and $\phi$ denotes two different scalar particles with the first label in the parentheses being the masses of the two, second being the $4$-momentum and the third denotes the direction in color space. The $g$ is a gluon which is labelled by the polarisation, $4$-momentum and the color index.\\
From the classical radiation perspective the frequency of the radiation is very low as compared to the characteristic energy of the system given by the inverse of the impact parameter.  This means that all the gluons are emitted in the same frequency bin and hence, one needs to consider the simultaneous limit of the multiple soft gluon theorem. 
Such simultaneous limits have a rather intricate structure \cite{berends} and hence we re-derive it in some detail below.\\
We start with the double soft gluon theorem at leading order and then give a general derivation for the multiple soft gluon theorem at leading order.

\subsubsection{Leading order expansion}

In $D\  \textgreater \ 4$ dimensions, the leading soft gluon theorem is a universal factorization theorem. It is independent of the details of the scattering states and does not receive any loop corrections.

We thus consider an amplitude consisting of two soft gluons, four scalar particles and $N-2$ hard gluons. The convention followed is that the first two scalar particles are taken to be incoming and the rest two to be outgoing. Hence, the values of $\eta$ are
\begin{align}
 \eta_{i} = -1,\ \ \ \ \ \ \ i=1,2 \\
          = +1 \ \ \ \ \ \ \ i=3,4.
\end{align}
The soft and the hard gluons are taken to be outgoing.

 Then the amplitude is 
\begin{multline}
 \mathcal{A}^{a_{1},a_{2},a_{3},a_{4},b_{1},...b_{N}}_{N+4}(1,2,3,4;k_{1},k_{2},...k_{N}) = \\
 \sum_{i=1}^{4} \eta_{i}f^{b_{1}a_{i}c_{i}}S_{i}(\epsilon_{1},k_{1})\bigg[\sum_{\substack{j=1  \\ j\neq i}}^{4} \eta_{j}f^{b_{2}a_{j}c_{j}}S_{j}(\epsilon_{2},k_{2})\mathcal{A}^{a_{1}..c_{i},..c_{j}..a_{4},b_{3},...b_{N}}_{N+2}(1,2,3,4;k_{3},...k_{N})\\
 +\sum_{m=3}^{N}f^{b_{2}b_{m}c_{m}}S_{m}(\epsilon_{2},k_{2})\mathcal{A}^{a_{1},..c_{i}..,a_{4},b_{3},.c_{m}..b_{N}}_{N+2}(1,2,3,4;k_{3},...k_{N})\bigg]  \\ 
  + \sum_{i=1}^{4} f^{b_{1}a_{i}c}S_{i}(\epsilon_{1},k_{1})f^{b_{2}cd_{i}}\frac{\epsilon_{2}\cdot p_{i}}{p_{i}\cdot(k_{1}+k_{2})} \mathcal{A}^{a_{1},..,d_{i},..,a_{4},b_{3},..,b_{N}}_{N+2}(1,2,3,4;k_{3},...k_{N})\\
 +\sum_{i=3}^{N}f^{b_{1}b_{i}c_{i}}S_{i}(\epsilon_{1},k_{1})\bigg[\sum_{j=1}^{4} \eta_{j}f^{b_{2}a_{j}c_{j}}S_{j}(\epsilon_{2},k_{2})\mathcal{A}^{a_{1}..c_{j}..a_{4},b_{3},.c_{i}..b_{N}}_{N+2}(1,2,3,4;k_{3},...k_{N})\\
 +\sum_{\substack{m=3 \\ m \neq i}}^{N}f^{b_{2}b_{m}c_{m}}S_{m}(\epsilon_{2},k_{2})\mathcal{A}^{a_{1},..,a_{4},b_{3},.c_{i}..c_{m}..b_{N}}_{N+2}(1,2,3,4;k_{3},...k_{N})\bigg]\\
 + \sum_{i=3}^{N} f^{b_{1}b_{i}c}S_{i}(\epsilon_{1},k_{1})f^{b_{2}cd_{i}}\frac{\epsilon_{2}\cdot k_{i}}{k_{i}\cdot(k_{1}+k_{2})} \mathcal{A}^{a_{1},..,a_{4},b_{1},..,d_{i}..,b_{N}}_{N+2}(1,2,3,4;k_{3},...k_{N})\\
 +f^{b_{1}b_{2}c_{2}}S_{2}(\epsilon_{1},k_{1})\bigg[\sum_{j=1}^{4} \eta_{j}f^{c_{2}a_{j}d_{j}}\frac{\epsilon_{2}\cdot p_{j}}{p_{j}\cdot (k_{1}+k_{2})}\mathcal{A}^{a_{1}..d_{j}..a_{4},b_{3},..,b_{N}}_{N+2}(1,2,3,4;k_{3},...k_{N})\\
 +\sum_{m=3}^{N}f^{c_{2}b_{m}c_{m}}\frac{\epsilon_{2}\cdot k_{m}}{k_{m}\cdot (k_{1}+k_{2})}\mathcal{A}^{a_{1},..,a_{4},b_{3},..c_{m}..b_{N}}_{N+2}(1,2,3,4;k_{3},...k_{N})\bigg] + (g_{1} \leftrightarrow g_{2}).
\end{multline}
The color indices of the scalar particles gets rotated in color space by the action of the Lie algebra generators (the soft gluons). Also, 
\begin{equation}
 S_{i}(\epsilon_{1},k_{1}) = S^{(0)}_{i}(\epsilon_{1},k_{1})
\end{equation}
The expression for which is 
\begin{equation}
 S_{i}^{(0)}(\epsilon,k) = \frac{\epsilon\cdot p_{i}}{p_{i}\cdot k}. 
\end{equation}
The $p_{i}$s are the momenta of the external hard particles.\\
This expression can schematically be written in a slightly more condensed form 
\begin{multline} \label{eq:4}
\mathcal{A}^{a_{1},a_{2},a_{3},a_{4},b_{1},...b_{N}}_{N+4}(1,2,3,4;k_{1},k_{2},...k_{N}) = \\
 \bigg[\sum_{i=1}^{4} \eta_{i}T^{b_{1}}_{i}S_{i}(\epsilon_{1},k_{1})\bigg]\bigg[\sum_{\substack{j=1 \\ j\neq i}}^{4} \eta_{j}T^{b_{2}}_{j}S_{j}(\epsilon_{2},k_{2})\bigg]\mathcal{A}^{a_{1}...a_{4},b_{3},...b_{N}}_{N+2}(1,2,3,4;k_{3},...k_{N})\\
 + \sum_{i=1}^{4} T^{b_{1}}_{i}S_{i}(\epsilon_{1},k_{1})T^{b_{2}}_{i}\frac{\epsilon_{2}\cdot p_{i}}{p_{i}\cdot (k_{1}+k_{2})} \mathcal{A}^{a_{1},...,a_{4},b_{3},..,b_{N}}_{N+2}(1,2,3,4;k_{3},...k_{N})\\
 +\bigg[\sum_{i=1}^{4} \eta_{i}T^{b_{1}}_{i}S_{i}(\epsilon_{1},k_{1})\bigg]\sum_{m=3}^{N}T^{b_{2}}_{m}S_{m}(\epsilon_{2},k_{2})\mathcal{A}^{a_{1},..a_{4},b_{3}..b_{N}}_{N+2}(1,2,3,4;k_{3},...k_{N})\\
 +\sum_{i=3}^{N}T^{b_{1}}_{i}S_{i}(\epsilon_{1},k_{1})\bigg[\sum_{j=1}^{4} \eta_{j}T^{b_{2}}_{j}S_{j}(\epsilon_{2},k_{2})\bigg]\mathcal{A}^{a_{1}....a_{4},b_{3},...b_{N}}_{N+2}(1,2,3,4;k_{3},...k_{N})\\
 +\sum_{i=3}^{N}T^{b_{1}}_{i}S_{i}(\epsilon_{1},k_{1})\sum_{\substack{m=3 \\ m\neq i}}^{N}T^{b_{2}}_{m}S_{m}(\epsilon_{2},k_{2})\mathcal{A}^{a_{1},..,a_{4},b_{3},...b_{N}}_{N+2}(1,2,3,4;k_{3},...k_{N})\\
 +\sum_{i=3}^{N} T^{b_{1}}_{i}S_{i}(\epsilon_{1},k_{1})T^{b_{2}}_{i}\frac{\epsilon_{2}\cdot k_{i}}{k_{i}\cdot (k_{1}+k_{2})} \mathcal{A}^{a_{1},...,a_{4},b_{1},...,b_{N}}_{N+2}(1,2,3,4;k_{3},...k_{N})\\
 +f^{b_{1}b_{2}c_{2}}S_{2}(\epsilon_{1},k_{1})\bigg[\sum_{j=1}^{4} \eta_{j}T^{c_{2}}_{j} \frac{\epsilon_{2}\cdot p_{j}}{p_{j}\cdot (k_{1}+k_{2})}\bigg]\mathcal{A}^{a_{1}...a_{4},b_{3},..,b_{N}}_{N+2}(1,2,3,4;k_{3},...k_{N})\\
 +f^{b_{1}b_{2}c_{2}}S_{2}(\epsilon_{1},k_{1})\sum_{m=3}^{N}T^{c_{2}}_{m}\frac{\epsilon_{2}\cdot k_{m}}{k_{m}\cdot (k_{1}+k_{2})}\mathcal{A}^{a_{1},..,a_{4},b_{3},...b_{N}}_{N+2}(1,2,3,4;k_{3},...k_{N})\\ + (g_{1} \leftrightarrow g_{2}) + \mathcal{O}(1/\omega).
\end{multline}
$T^{b_{i}}_{j}$ denotes the soft gluon acting on an external hard particle. The last two contributions come when one soft gluon is emitted from the other soft gluon. Such terms are collectively called contact terms.\\
The symbol $(g_{1} \leftrightarrow g_{2})$ signifies that one has to sum over both the orderings of the soft gluons. It is to be noted that the terms which have to be summed over are the ones in which the two soft gluons are emitted from the same external leg.\\ In writing the expression eq.\eqref{eq:4}, we have dropped terms which are subleading in the soft energy, $\mathcal{O}(1/\omega)$. The class of diagrams which contribute at the subleading order are 
\begin{itemize}
\item The first comes from the soft expansion of diagrams with the soft gluons being emitted from the external legs. The subleading term comes from the expansion of the propogator. 
\item The second includes diagrams with $4$-point vertices, which includes contributions from the vertex in YM and scalars coupled to YM. For example when the two soft gluons are emitted from the same scalar particle, the expression for the diagram is
\begin{equation}
 \sum_{i=1}^{4} f^{a_{i}b_{1}c}f^{cb_{2}d_{i}}\frac{(\epsilon_{1}\cdot \epsilon_{2})}{2p_{i}\cdot(k_{1}+k_{2})}\mathcal{A}_{N+2}^{a_{1}..d_{i}..a_{4}b_{3}...b_{N}}(1,2,3,4;k_{3},..,k_{N}). 
\end{equation}
From the denominator one can see that the term is of order $\mathcal{O}(1/\omega)$, which is subleading for double soft gluon emissions. It can be seen that diagrams which involve the YM $4$-point vertices also give the same order of magnitude.
\item The last class of diagrams are those where the soft gluons are emitted from the propagators. Such terms start at the subleading order in soft energy.
\end{itemize}
Eq.\eqref{eq:4} can be further condensed by writing schematically $f$ for a contact term and $T$ for the emission of the soft gluon from any of the hard particles. In this form eq.\eqref{eq:4} can be written as
\begin{equation} \label{eq:51}
 \mathcal{A}_{4+N} = (TT + fT+Tf)\  \mathcal{A}_{4+N-2}.
\end{equation}
In writing the above expression everything apart from the structure constant and the Lie algebra generator has been supressed to emphasize the type of particle from which the two soft gluons are emitted. It can be seen that all the terms in eq.\eqref{eq:4} can be written like one of the terms in eq.\eqref{eq:51}.\\
As we will show below \eqref{eq:7}, the expression in eq.\eqref{eq:4} simplifies drastically when the colored massive scattering states are taken to be classical (e.g. composed of large number of quarks), namely they drop out. Hence, anticipating this simplification we collect all the contact terms in one group and refer to it as ${\cal R}$. 
The vanishing of contact terms at leading order turns out to be true irrespective of color configuration of soft gluons.\footnote{For monochromatic radiation, that is, if all soft gluons are emitted in the same bin, contact terms again drop out, but due to the anti-symmetry of $f^{abc}$. Hence in that case, the multiple soft gluon theorem simplifies drastically even when scattering states are composed of only a small number of quarks.} 
So, in this case, the expression becomes
\begin{multline}
 \mathcal{A}^{a_{1},a_{2},a_{3},a_{4},b_{1},b_{2},b_{3}...b_{N}}_{4+N}(1,2,3,4;k_{1},k_{2},...k_{N}) = \\
 \bigg[\sum_{i=1}^{4} \eta_{i}T^{b_1}_{i}S_{i}(\epsilon_{1},k_{1})\bigg]\bigg[\sum_{\substack{j=1 \\ j\neq i}}^{4} \eta_{j}T^{b_2}_{j}S_{j}(\epsilon_{2},k_{2})\bigg]\mathcal{A}^{a_{1}...a_{4},b_{3},...b_{N}}_{N+2}(1,2,3,4;k_{3},...k_{N})\\
  +\bigg[\sum_{i=1}^{4} \eta_{i}T^{b_{1}}_{i}S_{i}(\epsilon_{1},k_{1})\bigg]\sum_{m=3}^{N}T^{b_{2}}_{m}S_{m}(\epsilon_{2},k_{2})\mathcal{A}^{a_{1},..a_{4},b_{3}..b_{N}}_{N+2}(1,2,3,4;k_{3},...k_{N})\\
 +\sum_{i=3}^{N}T^{b_{1}}_{i}S_{i}(\epsilon_{1},k_{1})\bigg[\sum_{j=1}^{4} \eta_{j}T^{b_{2}}_{j}S_{j}(\epsilon_{2},k_{2})\bigg]\mathcal{A}^{a_{1}....a_{4},b_{3},...b_{N}}_{N+2}(1,2,3,4;k_{3},...k_{N})\\
 +\sum_{i=3}^{N}T^{b_{1}}_{i}S_{i}(\epsilon_{1},k_{1})\sum_{\substack{m=3 \\ m\neq i}}^{N}T^{b_{2}}_{m}S_{m}(\epsilon_{2},k_{2})\mathcal{A}^{a_{1},..,a_{4},b_{3},...b_{N}}_{N+2}(1,2,3,4;k_{3},...k_{N})\\
 +\sum_{i=3}^{N} T^{b_{1}}_{i}S_{i}(\epsilon_{1},k_{1})T^{b_{2}}_{i}\frac{\epsilon_{2}\cdot k_{i}}{k_{i}\cdot (k_{1}+k_{2})} \mathcal{A}^{a_{1},...,a_{4},b_{3},...,b_{N}}_{N+2}(1,2,3,4;k_{3},...k_{N})\\
 + \sum_{i=1}^{4} T^{b_{1}}_{i}S_{i}(\epsilon_{1},k_{1})T^{b_{2}}_{i}\frac{\epsilon_{2}\cdot p_{i}}{p_{i}\cdot (k_{1}+k_{2})} \mathcal{A}^{a_{1},...,a_{4},b_{3},..,b_{N}}_{N+2}(1,2,3,4;k_{3},...k_{N})\\
 +\sum_{i=3}^{N} T^{b_{2}}_{i}S_{i}(\epsilon_{2},k_{2})T^{b_{1}}_{i}\frac{\epsilon_{1}\cdot k_{i}}{k_{i}\cdot (k_{1}+k_{2})} \mathcal{A}^{a_{1},...,a_{4},b_{1},...,b_{N}}_{N+2}(1,2,3,4;k_{3},...k_{N})\\
 + \sum_{i=1}^{4} T^{b_{2}}_{i}S_{i}(\epsilon_{2},k_{2})T^{b_{1}}_{i}\frac{\epsilon_{1}\cdot p_{i}}{p_{i}\cdot (k_{1}+k_{2})} \mathcal{A}^{a_{1},...,a_{4},b_{3},..,b_{N}}_{N+2}(1,2,3,4;k_{3},...k_{N})\\
\hspace*{-1.9in} +\ {\cal R}\ \mathcal{A}^{a_{1},...,a_{4},b_{3},..,b_{N}}_{N+2}(1,2,3,4;k_{3},...k_{N}).
\end{multline}
The last two terms before the contact terms ($\mathcal{R}$) come from summing over both ordering of the gluons. These terms differ from the fifth and sixth terms in the above equation by a contact term as can be seen by using the commutation relation,
\begin{equation}
 T^{b_{1}}T^{b_{2}} - T^{b_{2}}T^{b_{1}} = if^{b_{1}b_{2}c}T^{c}.
\end{equation}
Hence, one can switch the orderings in the last two terms and the difference can be put into the contact terms ($\mathcal{R}$). Then the amplitude can be written compactly as
\begin{multline}
 \mathcal{A}^{a_{1},a_{2},a_{3},a_{4},b_{1},b_{2},b_{3}...b_{N}}_{4+N}(1,2,3,4;k_{1},k_{2},...k_{N})= \\ 
     \bigg[ g \sum_{i_{1}=1}^{4} \eta_{i_1}T^{b_{1}}_{i_1}S_{i_1}(\epsilon_{1},k_{1})\bigg]\bigg[ g \sum_{i_{2}=1}^{4} \eta_{i_2}T^{b_{2}}_{i_{2}}S_{i_2}(\epsilon_{2},k_{2})\bigg] 
              \mathcal{A}^{a_{1},a_{2},a_{3},a_{4},b_{3},...b_{N}}_{4+N-2} (1,2,3,4;k_{3},...k_{N})\\
              + \bigg[ g \sum_{i_{1}=3}^{N} T^{b_{1}}_{i_1}S_{i_1}(\epsilon_{1},k_{1})\bigg]\bigg[ g \sum_{i_{2}=1}^{N} T^{b_{2}}_{i_{2}}S_{i_2}(\epsilon_{2},k_{2})\bigg] 
              \mathcal{A}^{a_{1},a_{2},a_{3},a_{4},b_{3},...b_{N}}_{4+N-2} (1,2,3,4;k_{3},...k_{N})\\
+\ {\cal R}\ \mathcal{A}^{a_{1},a_{2},a_{3},a_{4},b_{3},...b_{N}}_{4+N-2} (1,2,3,4;k_{3},...k_{N}).           
\end{multline}
This is the double soft gluon expansion with terms upto the leading order.\\
In the case of $N$ soft gluons the leading order result turns out to be  
\begin{multline}
\label{eq:6}
 \mathcal{A}^{a_{1},a_{2},a_{3},a_{4},b_{1},...,b_{N}}_{4+N}(1,2,3,4;k_{1},k_{2},...k_{N}) = \\ 
     \bigg[ g \sum_{i_{1}=1}^{4} \eta_{i_1}T^{b_{1}}_{i_1}S_{i_1}(\epsilon_{1},k_{1})\bigg]...\bigg[ g \sum_{i_{N}=1}^{4} \eta_{i_N}T^{b_{N}}_{i_N}S_{i_N}(\epsilon_{N},k_{N})\bigg] 
              \mathcal{A}^{a_{1},a_{2},a_{3},a_{4}}_{4} (1,2,3,4)\\ + \mathcal{R}\ \mathcal{A}^{a_{1},a_{2},a_{3},a_{4}}_{4} (1,2,3,4) 
        \end{multline}
The $\mathcal{A}^{a_{1},a_{2},a_{3},a_{4}}_{4} (1,2,3,4)$ is the amplitude which doesn't have any soft gluon and is the $4$-point amplitude of scattering the two massive scalars.\\
We start with the expression for the multiple soft gluon theorem at the leading order. For this case the number of graphs which will contribute to the leading order, $\mathcal{O}(1/\omega^{N})$, will be extremely high. 
The amplitude can be written schematically like eq.\eqref{eq:51} by enumerating all the possible ways of arranging $f$s and $T$s in an $N$-dimensional array but with the last entry being a $T$. This can be written as
\begin{equation} 
 \mathcal{A}_{4+N} = ( TTT...T +fTT...T+...+fff....fT)\  \mathcal{A}_{4}.\label{eq:56}
\end{equation}
There are $2^{N-1}$ terms in this expression and all the terms are contact terms except the first term.\\
Assuming that contact terms drop out, we are left with only the first term.

\subsubsection{Subleading order expansion}

Our analysis in the previous section was restricted to the derivation of multiple soft gluon theorem at leading order ($\mathcal{O}(\frac{1}{\omega^{N}})$) in soft frequency. As we saw, even at leading order, the multiple soft gluon theorem is a rather intricate beast with contact terms arising due to color degrees of freedom that gluons possess. But as our interest lies in understanding the scattering of classical colored particles, all the contact terms drop out, the soft theorem ``abelianizes" and the result is analogous to the multiple soft graviton theorem at leading order.\\ 
We would now like to extend the analysis to subleading order in soft frequency, i.e. to order $\mathcal{O}(\frac{1}{\omega^{N-1}})$. The subleading soft gluon factor is sensitive to infra-red divergences in  $D \leq\ 5$ dimensions and hence we restrict our analysis to $D\ >\ 5$ dimensions. However even in this case, the subleading soft factor receives correction from quantum loop effects \cite{biswajit} as well as from a certain class of irrelevant operators \cite{elvang}. More in detail,

\begin{equation}
S^{(1)a}(\epsilon, k)\ =\ S^{(1)a}_{\textrm{universal}}(\epsilon, k)\ +\ \tilde{S}^{(1) a}(\epsilon,k),
\end{equation}
where the universal soft factor $S^{(1) a}_{\textrm{universal}}(\epsilon, k)$ is given by 
\begin{equation}
S^{(1) a}_{\textrm{universal}}(\epsilon, k)\ =\ \sum_{i=1}^{n}\frac{\epsilon_{\mu}{\bf J}_{i}^{\mu\nu}k_{\nu}}{p_{i}\cdot k}T^{a}_{i}.
\end{equation}
Here ${\bf{J}}_{i}$ is the total angular momentum of the $i^{th}$ particle. As the scattering processes we are concerned with have impact paramater much larger then the size of the objects, spin angular momentum is sub-dominant (in $b^{-1}$) as compared to the orbital angular momentum and can be ignored. Moreover, all the non-universal terms which come from infra-red finite loop corrections depend on the internal structure of the scatterer \cite{ashoke} and are also sub-dominant compared to the universal term. Thus  we only consider the universal part of the subleading soft gluon factor and may also ignore the contribution of the spin angular momentum if the scattering objects have spin.\\
But there is an additional subtlety. A general analysis of multiple soft gluon theorem at subleading order seems to be unavailable in the literature. This is true even if we consider only the universal term in the subleading soft factor for the single gluon emission. We make progress by the lesson learnt from the previous section. By considering the example of double soft gluon theorem, it is not difficult to see that the complications arise due to contact terms which have two types of origins. One due to the fact that a soft gluon can emit yet another soft gluon and the other is due to presence of $4$-point vertices which couple colored scalars to two gluons.  However due to the fact that we are interested in radiation emitted by classical colored particles all the contact terms drop out even at subleading order and our analysis can be extended. In effect the multiple soft gluon theorem then reduces to considering soft emission in the consecutive limit.\\
Again let us consider the scattering process of two soft gluons, four scalar particles and $N-2$ hard gluons. Keeping terms till the subleading order in soft expansion, the amplitude in the consecutive limit is 
\begin{multline} \label{eq:136}
\mathcal{A}^{a_{1},a_{2},a_{3},a_{4},b_{1},...b_{N}}_{N+4}(1,2,3,4;k_{1},k_{2},...k_{N}) = \\
 \bigg[\sum_{i=1}^{4} \eta_{i}T^{b_{1}}_{i}S_{i}(\epsilon_{1},k_{1})\bigg]\bigg[\sum_{\substack{j=1 \\ j\neq i}}^{4} \eta_{j}T^{b_{2}}_{j}S_{j}(\epsilon_{2},k_{2})\bigg]\mathcal{A}^{a_{1}...a_{4},b_{3},...b_{N}}_{N+2}(1,2,3,4;k_{3},...k_{N})\\
 +\bigg[\sum_{i=1}^{4} \eta_{i}T^{b_{1}}_{i}S_{i}(\epsilon_{1},k_{1})\bigg]\sum_{m=3}^{N}T^{b_{2}}_{m}S_{m}(\epsilon_{2},k_{2})\mathcal{A}^{a_{1},..a_{4},b_{3}..b_{N}}_{N+2}(1,2,3,4;k_{3},...k_{N})\\
 +\sum_{i=3}^{N}T^{b_{1}}_{i}S_{i}(\epsilon_{1},k_{1})\bigg[\sum_{j=1}^{4} \eta_{j}T^{b_{2}}_{j}S_{j}(\epsilon_{2},k_{2})\bigg]\mathcal{A}^{a_{1}....a_{4},b_{3},...b_{N}}_{N+2}(1,2,3,4;k_{3},...k_{N})\\
 +\sum_{i=3}^{N}T^{b_{1}}_{i}S_{i}(\epsilon_{1},k_{1})\sum_{\substack{m=3 \\ m\neq i}}^{N}T^{b_{2}}_{m}S_{m}(\epsilon_{2},k_{2})\mathcal{A}^{a_{1},..,a_{4},b_{3},...b_{N}}_{N+2}(1,2,3,4;k_{3},...k_{N})\\
 +f^{b_{1}b_{2}c_{2}}S_{2}(\epsilon_{1},k_{1})\bigg[\sum_{j=1}^{4} \eta_{j}T^{c_{2}}_{j} S_{j}(\epsilon_{2},k_{2})\bigg]\mathcal{A}^{a_{1}...a_{4},b_{3},..,b_{N}}_{N+2}(1,2,3,4;k_{3},...k_{N})\\
 +f^{b_{1}b_{2}c_{2}}S_{2}(\epsilon_{1},k_{1})\bigg[\sum_{m=3}^{N}T^{c_{2}}_{m}S_{m}(\epsilon_{2},k_{2})\bigg]\mathcal{A}^{a_{1},..,a_{4},b_{3},...b_{N}}_{N+2}(1,2,3,4;k_{3},...k_{N}) + \mathcal{O}(1).
\end{multline}
Here  
\begin{equation} \label{eq:110}
 S_{i}(\epsilon,k) = \frac{\epsilon\cdot p_{i}}{p_{i}\cdot k} - \frac{\epsilon_{\mu}k_{\nu}J^{\mu\nu}_{i}}{p_{i}\cdot k} ,\ \ \text{for}\ \ \ \ \ \ i=1,2
\end{equation}
and 
\begin{equation} \label{eq:116}
 S_{i}(\epsilon,k) = \frac{\epsilon\cdot p_{i}}{p_{i}\cdot k} + \frac{\epsilon_{\mu}k_{\nu}J^{\mu\nu}_{i}}{p_{i}\cdot k} ,\ \ \text{for}\ \ \ \ \ \ i=3,4.
\end{equation}
For the hard gluons the $S_{i}(\epsilon,k)$ is given by the last equation as they are all outgoing.	
The angular momentum operator $J_{i}^{\mu\nu}$ is
\begin{equation} \label{eq:111}
 J^{\mu\nu}_{i} = p_{i}^{\mu}\frac{\partial}{\partial p_{i\nu}} - p_{i}^{\nu}\frac{\partial}{\partial p_{i\mu}}.
\end{equation} 
In eq.\eqref{eq:136} the last two terms are the contact terms.
These terms will drop out when the scalar particles are taken to be classical colored particles.\\
Hence, the double soft gluon theorem in the consecutive limit is
\begin{multline}
 \mathcal{A}^{a_{1},a_{2},a_{3},a_{4},b_{1},b_{2},b_{3}...b_{N}}_{4+N}(1,2,3,4;k_{1},k_{2},...k_{N})= \\ 
     \bigg[ g \sum_{i_{1}=1}^{4} \eta_{i_1}T^{b_{1}}_{i_1}S_{i_1}(\epsilon_{1},k_{1})\bigg]\bigg[ g \sum_{i_{2}=1}^{4} \eta_{i_2}T^{b_{2}}_{i_{2}}S_{i_2}(\epsilon_{2},k_{2})\bigg] 
              \mathcal{A}^{a_{1},a_{2},a_{3},a_{4},b_{3},...b_{N}}_{4+N-2} (1,2,3,4;k_{3},...k_{N})\\
              + \bigg[ g \sum_{i_{1}=3}^{N} T^{b_{1}}_{i_1}S_{i_1}(\epsilon_{1},k_{1})\bigg]\bigg[ g \sum_{i_{2}=1}^{N} T^{b_{2}}_{i_{2}}S_{i_2}(\epsilon_{2},k_{2})\bigg] 
              \mathcal{A}^{a_{1},a_{2},a_{3},a_{4},b_{3},...b_{N}}_{4+N-2} (1,2,3,4;k_{3},...k_{N}) + \mathcal{O}(k)\\ + \mathcal{R}\ \mathcal{A}^{a_{1},a_{2},a_{3},a_{4},b_{3},...b_{N}}_{4+N-2} (1,2,3,4;k_{3},...k_{N}).
\end{multline}
The multiple soft gluon case follows the same logic as above, namely all the contact terms will drop out for classical colored particles. Hence, the multiple soft subleading expansion is
\begin{multline} \label{eq:109}
 \mathcal{A}^{a_{1},a_{2},a_{3},a_{4},b_{1},...b_{N}}_{4+N}(1,2,3,4;k_{1},k_{2},...k_{N}) = \\ 
     \bigg[ g \sum_{i_{1}=1}^{4} \eta_{i_1}T^{b_{1}}_{i_1}S_{i_1}(\epsilon_{1},k_{1})\bigg]...\bigg[ g \sum_{i_{N}=1}^{4} \eta_{i_N}T^{b_{N}}_{i_N}S_{i_N}(\epsilon_{N},k_{N})\bigg] 
              \mathcal{A}^{a_{1},a_{2},a_{3},a_{4}}_{4} (1,2,3,4)\\ + \mathcal{R}\ \mathcal{A}^{a_{1},a_{2},a_{3},a_{4}}_{4} (1,2,3,4) .
\end{multline}
with the $S_{i}(\epsilon,k)$s being given by eq.\eqref{eq:110} and eq.\eqref{eq:116}.
This is the amplitude we will use to compute the low frequency gluon radiation. The above amplitude is written schematically which makes it look simple, but once the action of the Lie algebra generator on the color spaces on the scalar particles is taken into account, the expression for the amplitude becomes complitcated.\\
The action of the gluon generator is
\begin{equation}
 T^{a}_{A} \ket{b_{i}} = \sum_{c=1}^{N^{2}-1}-if^{b_{i}ac} \ket{c}.
\end{equation}
The equation is written for the $i^{th}$ scalar particle which lives in the adjoint representation of the gauge group $SU(N)$.
The matrix valuedness of the single soft gluon factor inhibits a straight forward saddle point analysis as compared to the soft graviton case. But as we will see in the next section, for classical colored particles the amplitude will become quite simple.   

\subsection{Computing soft radiation from soft theorems $:$\\ The McLerran Venugopalan model} \label{Extremisation}

In this section we show that just as in the case of gravitational scattering, we  can use eqn.\eqref{eq:109} to compute low frequency gluon radiation from the scattering of classical objects with color. For simplicity we take the scattering objects to be scalars with no spin, however our analysis can be generalised to the spinning case as well.\\
A new subtlety arises when analysing low frequency gluon radiation from scattering amplitudes, which was not present in gravity. In order to compute classical radiation from the soft gluon theorem, we need to analyse the asymptotic data of the scattering states.  As in the case of gravity, where classical radiation emitted from large composite objects could be computed using soft theorems, here we consider colored states composed of large number of elementary particles (such as quarks) and which possess non-trivial color charges.\footnote{An immediate example that comes to mind is collision of heavy Nuclei.} \footnote{Here our interpretation of the classical scattering differs from that of \cite{goldberger} where the classical particles were considered to be semi-classical single-particle states. Our proof of classical double copy in the soft limit will however be applicable to the results of \cite{goldberger} as well.}\\
How does one model color degree of freedom of such objects which are composed of large number of elementary colored particles such as quarks? A beautiful answer to this question was provided by Mclerran, Venugopalan and Jeon in a series of papers \cite{mcLerran,venu}. In essence, they showed that the color of an object  composed of $k$ number of quarks is peaked sharply around a representation of weight of  high order, namely $\sqrt{k}$. Using this result, they showed that one can indeed model a colored object consisting of large number of quarks by ``classical charge" $c^{a}$ such that $\sum_{a}c^{a}c_{a}\ =\ C_{k}$ where $C_{k}$ is the quadratic Casimir of the gauge group in representation $k$. 
An interesting off-shoot of this analysis is that an action of a single  color generator has negligible effect on color as $k\ >>\ 1$. Quantitatively this implies that one is in a coherent state in the color space,\footnote{In fact in \cite{venu}, the authors showed that their construction of replacing color degrees of freedom of a composite object by a classical color can indeed be thought of as a coherent state representation in the color space.} 

\begin{equation}
\label{eq:7}
T^{a}\vert c\rangle = c^{a} \vert c\rangle,
\end{equation}
$T^{a}$ is the Lie algebra generator and $c^{a}$ is the component of the vector in the color space. Hence we have a definition of classical color degree of freedom, as a label of a coherent state in the color space. 
Hence when the scattering states involved in the (quantum) scattering amplitude are classical then when a single gluon is emitted, we can substitute $T^{a}_{i}\rightarrow c^{a}_{i}$. Note that for a ``macroscopic object" in which a large number of soft gluons are emitted, this replacement may break down unless the number of emitted gluons are far smaller then the number $\sqrt{k}$. For now, we assume this to be the case. Below equation \eqref{eq:140} we will argue that this assumption is self-consistent. 

An immediate consequence of using the MV model for defining classical color charge is the following. As $c^{a}\ \sim\ O(\sqrt{k})\ >>\ 1$ (in appropriate units) the contact terms (which we denoted as ${\cal R}$ in the previous section) are subdominant compared to the leading term in the multiple soft gluon theorem. This is because ratio of the contact terms to the leading term goes as $\frac{1}{\sqrt{k}}$. This justifies the assumption we made in deriving \eqref{eq:109}.\\
Thus when scattering states are in a color coherent state, the single soft leading gluon theorem for classical particles is,
\begin{multline}
  \lim_{\omega\rightarrow 0} \omega\mathcal{A}^{a}(c_{1},c_{2},c_{3},c_{4};k) =  \\
 g\bigg[- c^{a}_{1}\frac{\epsilon\cdot p_{1}}{p_{1}\cdot \hat{k}} - c^{a}_{2}\frac{\epsilon\cdot p_{2}}{p_{2}\cdot \hat{k}} + c^{a}_{3}\frac{\epsilon\cdot p_{3}}{p_{3}\cdot \hat{k}}  
    + c^{a}_{4}\frac{\epsilon\cdot p_{4}}{p_{4}\cdot \hat{k}}\bigg] \mathcal{A}(c_{1},c_{2},c_{3},c_{4}) + \mathcal{O}(1).  
\end{multline}
Here $k = \omega\hat{k} $. Note that this looks very similar to the factorization which occurs in QED. The only difference is that the electric charges are replaced with (vector) color charges of the respective states.  \\
We will now use the multiple soft gluon theorem to  calculate the flux emitted in soft frequency domain from the scattering of two colored particles. The class of scattering processes we consider are the same as the ones considered in \cite{goldberger}. This class can be parametrized by placing the following constraint on the color charges of the initial states and the impact parameter, 
\begin{equation}
g^{2}\frac{c_{1}\cdot c_{2}}{Eb^{D-3}} < < 1.      
\end{equation}
Here $E \sim M$ is the energy of the colored particles, $b$ is the impact paramter and $c_{1}$ and $c_{2}$ are the initial color charges of the two different classical colored particles. The above relation is obtained by requiring that the deflection in the path of the classical colored particles is very much less than the impact parameter. This has to be done at each order in perturbation theory, with the above quantity being the relevant expansion parameter.\\ 
Consider the amplitude for the emission of $N$ soft gluons from the scattering of two scalar classical particles, 
\begin{multline} \label{eq:80}
 \Phi(p_{1},c_{1},M) + \phi(p_{2},c_{2},m) \rightarrow \Phi(p_{3},c_{3},M) + \phi(p_{4},c_{4},m) \\ 
 + g(\epsilon_{1},k_{1},b_{1}) + g(\epsilon_{2},k_{2},b_{2})+...+ g(\epsilon_{n},k_{n},b_{n}).
\end{multline}
The soft gluons are all in the same bin while the colored scalars have color charges $c_{i}$s and momentum $p_{i}$. Since the external legs are classical particles we can replacing the $T^{a}_{i}$ with $c^{a}_{i}$ in equation \eqref{eq:109} and we get
\begin{multline}
\label{eq:10}
 \mathcal{A}_{4+N}(c_{1},c_{2},c_{3},c_{4};k_{1},k_{2},...k_{N})= \\ 
    \bigg(g \sum_{i_{1}=1}^{4}\eta_{i_{1}}c^{b_{1}}_{i_{1}}S_{i_{1}}(\epsilon_{1},k_{1})\bigg)...
    \bigg(g \sum_{i_{N}=1}^{4}\eta_{i_{N}}c^{b_{N}}_{i_{N}}S_{i_{N}}(\epsilon_{N},k_{N})\bigg)
              \mathcal{A}_{4} (c_{1},c_{2},c_{3},c_{4}) \\
              +\mathcal{O}\bigg(\frac{1}{\omega^{N-2}}\bigg).
              \end{multline}
As $c^{b}$s are just components of vectors in the color space unlike in equation \eqref{eq:6} where $T^{b}$s are operators, we can use the above factorisation relation to extract the low frequency gluon radiation from the soft theorem.\\
In the same vein the angular momentum operators at the subleading order are replaced by the (asymptotic) classical angular momenta defined as,
\begin{equation}
 J^{\mu\nu}_{i} = x^{\mu}_{i}p^{\nu}_{i} - x^{\nu}_{i}p^{\mu}_{i},
\end{equation}
where the $x_{i}$ is the asymptotic trajectory and $p_{i}$ is the asymptotic $4$-momentum of the classical colored particle.\\ 
We now calculate the total energy radiated out to infinity in a given bin and a color direction, $a$, from the above process. The differential cross section for the same is 
\begin{equation} \label{eq:60}
 d\sigma^{a} = \frac{1}{N!}(A^{a})^{N}|\mathcal{A}_{4} (c_{1},c_{2},c_{3},c_{4})|^{2}. 
\end{equation}
The $a$ index on the left hand side of the equation denotes that the differential cross section is for a particular direction in color space. The $N!$ is due to the indistinguishability of the soft gluons and $\mathcal{A}_{4} (c_{1},c_{2},c_{3},c_{4})$ is the hard amplitude without the soft gluons. The quantity $A^{a}$ has both the modulus square of the soft factor and the phase space factor 
\begin{equation}\label{eq:11}
 A^{a} = |g\sum_{i=1}^{4}\eta_{i}c^{a}_{i}S_{i}(\epsilon,k)|^{2} \frac{d^{D-1}\vec{k}}{(2\pi)^{D}2\omega}. 
\end{equation}
For massless particles using the dispersion relation $\omega = |\vec{k}|$, the phase space factor can be written as
\begin{equation}
 \frac{d^{D-1}k}{(2\pi)^{D}2\omega} = \frac{1}{2}\frac{\omega^{D-2}\delta\Delta\Omega}{(2\pi)^{D-1}}. 
\end{equation}
Plugging this back into eq.\eqref{eq:11},
\begin{equation}
\label{eq:14}
 A^{a} = \frac{1}{2}\frac{\omega^{D-2}\delta\Delta\Omega}{(2\pi)^{D-1}}|S^{a}(\epsilon,k)|^{2} \ \ \ ,\ \ S^{a}(\epsilon,k) =  g\sum_{i=1}^{4}\eta_{i}c^{a}_{i}S_{i}(\epsilon,k).
\end{equation}
The classical limit of the differential cross section is when $c^{2}>>1$ because from eq. \eqref{eq:14}, for a fixed value of $\omega$ and $\Delta\Omega$, $A^{a}(\epsilon,k) \sim |c^{a}|^{2}g^{2} \sim kg^{2}$ \footnote{There is no summation over $ a$ in this expression}. 
We shall now extremize the differential cross section by taking the logarithm of the same and differentiating with respect to $N$. We get
\begin{align}
 & \frac{\partial}{\partial N}(N \ln A^{a} - N \ln N + N) = 0 \label{eq:12}, \\ 
 & \therefore\  \ln N = \ln A^{a}.
\end{align}
Hence, the value of $N$ at which the probability distribution, in this bin, peaks is
\begin{equation} \label{eq:13}
 N_{0} = A^{a} = \frac{1}{2}\frac{\omega^{D-2}\delta\Delta\Omega}{(2\pi)^{D-1}}|S^{a}(\epsilon,k)|^{2}. 
\end{equation}
Since $N = A^{a}$, as follows from the reasoning below \eqref{eq:14}, in the classical limit $N$ is very large. 
Taking the double derivative of the function in \eqref{eq:12} we get
\begin{equation} \label{eq:140}
 \frac{\partial^{2}}{\partial N^{2}}(N \ln A^{a} - N \ln N + N)|_{N_{0}} = -\frac{1}{N_{0}},
\end{equation}
since $N_{0}$ is positive this shows that the probability distribution is indeed a maximum at this value of $N$.\\
As the number of soft gluons scale as $N_{0}\ \sim\ \vert c^{a}\vert^{2}\ g^{2}\omega^{D-4}\ \delta\triangle\Omega$. We see that for a fixed $\delta,\ \triangle\Omega$, $N_{0}\ \sim\ \frac{k\ g^{2}}{b^{D-4}}$. Whence given composite colored objects comprising of $k$ number of quarks, if the impact parameter is large enough, the number of soft gluons emitted is always negligible compared to the number of quarks. Whence to an excellent approximation we can assume that emission of $N_{0}$ soft gluons does not rotate the color vector $c_{i}$ of the scattering states.\\ 
One can also see that the differential cross section for this value of $N$ goes as $N^{N}/N!$ and for large $N$ this is a very large quantity.\\
Now that we have the expression for the number of soft gluons emitted, it is simple task to calculate the total energy flux radiated out to infinity in this bin which is obtained by $N_0$, in eq. \eqref{eq:13}, by the energy of a single soft gluon $\omega$
\begin{equation} 
 E^{a} = N_{0}\omega = \frac{1}{2}\frac{\omega^{D-1}\delta\Delta\Omega}{(2\pi)^{D-1}}|S^{a}(\epsilon,k)|^{2}. \label{eq:61}
\end{equation}
This is the total energy flux emitted in a given bin from the scattering of two massive colored classical particles. The important thing to note is that the flux is proportional to the modulus square of the single soft factor.\\ 
From this expression one can read off the radiative gauge field for the same process. The procedure is exactly analogous to the treatment for the gravity / electromagnetism, given in \cite{ashoke}. The result one gets is   
\begin{equation} \label{eq:130}
 \epsilon\cdot A^{a}(\omega,\vec{x}) = \mathcal{N}e^{i\omega R}S^{a}(\epsilon,k),\ \ \ \ \ |\vec{x}| = R. 
\end{equation}
Whence the low frequency gluon radiation for the classical scattering considered in \cite{goldberger} can be derived from multiple soft gluon theorem upto the subleading order in soft momentum.\\  
Now as $T^{a}$ is replaced by $c^{a}$ in the classical theory, we see that the classical double copy in soft expansion immediately follows from a ``double copy" construction of graviton soft factor from gluon soft factor. In essence this shows why the classical double copy works in our restricted setting. However we would like to give a first principle derivation of this statement. That is, starting from color-kinematics duality of the scattering amplitudes, we would like to see if the double copy construction of graviton soft factor $S_{GR}^{\mu \nu}(\epsilon,k)=S^{\mu \nu(0)}_{GR}(\epsilon,k)+S^{\mu\nu(1)}_{GR}(\epsilon,k)$
from the corresponding gluon soft factor is a consequence of the color-kinematics duality.

\section{Color Kinematics duality for soft theorems} \label{5}

As we saw in the previous section, the low frequency radiative gluon (gravitational) field in the classical scattering is proportional to single soft gluon (graviton) factor. Hence in order to derive the classical double copy, we need to show that there is a double copy at the level of soft factors and that this double copy is a consequence of the (quantum) color-kinematics duality\footnote{The double copy structure of soft factors has been noted in \cite{song,white,vera1}.}.\\
To do so we first need to start with an amplitude in theories with matter (massive / massless particles) coupled to YM and write it in a BCJ representation. For the case when the external particles are massless scalars, the duality was proved in \cite{vera}. A similar analysis for the case of massive colored scalars coupled to gluons was carried out in \cite{luna} and the corresponding amplitude was shown to satisfy the duality. In addition it was also shown in \cite{luna} that the double copied amplitude is an amplitude in dilaton gravity.\\ 
In addition to the progress done in \cite{vera,luna}, we will show that when the tree level amplitudes in scalar (massive) QCD are written in a manner such that the duality is manifest, the subleading soft gluon factor becomes ripe for the quantum double copy from which we get the subleading soft graviton factor. We refer to this procedure as \textit{soft double copy}.\\
A similar analysis has been carried out in \cite{oxburgh} in which it was shown that one can map the infra-red singularities of pure Yang-Mills theory to the ones in pure gravity by using the double copy for the eikonal integral factor, thereby providing evidence for the double copy to hold to all loops.\\ 
Our analysis can also be generalised to the BCJ representation of QCD amplitudes \cite{ochirov} although we anticipate an added subtlety in the analysis due to the intrinsic spin of the quarks. For quarks the angular momentum operator, in eq.\eqref{eq:111}, will have an extra term due the spin angular momentum of the quarks.
Due to this we restrict our analysis, in this work, to the scalar QCD case.\\  
Although our analysis is restricted to tree level scalar QCD, we believe that our results are more general. This is because the soft gluon and graviton factors are insensitive to loop corrections in $D\ >\ 5$ and hence we believe that our proof of soft double copy is valid even when loop corrections are included. There are impressive results regarding validity of color-kinematics duality and the douple copy at higher loop level in pure Yang-Mills theory \cite{bernloop}, but the proof of such a duality in theories of matter coupled to YM appears to be missing.\\
In the next section \ref{5-b}, we review the color-kinematics duality for tree level amplitudes in scalar QCD \cite{vera} and show that this duality leads to the double copy construction of the Weinberg soft graviton factor, $S^{\mu \nu(0)}_{GR}(\epsilon,k)$ from the leading soft gluon factor. In section \ref{5a}, we see if such a derivation can be extended to subleading order. That is, whether the color-kinematics duality also leads us to the double copy construction of subleading soft graviton factor from subleading soft gluon factor. However, we show that due to the fact that the subleading factor is an operator in momentum space, this is not the case.

\subsection{Leading soft gluon theorem}\label{5-b}

We now review the derivation of color-kinematics duality in scalar QCD \cite{vera} before using it to derive soft double copy relation.
Before dwelling into the details, let us summarise the key ideas involved in the proof. Although the proof is very general, we will consider a specific scattering of colored scalar particles and a single gluon.
\begin{itemize}
\item We start by writing a $5$-point tree level amplitude in scalar QCD and try to write it into the BCJ representation where the duality is manifest. This is done in a series of two steps by successive ``gauge transformations" of the kinematic factors.   
\item Once the amplitude is in the BCJ representation we expand the amplitude in the powers of the soft $4$-momentum. 
\item This leads us to the proof of the soft double copy, where the amplitude involving a single soft gluon can be transformed into an amplitude involving a single soft graviton by a double copy substitution obtained via BCJ relations. 
\end{itemize}
The process we consider is the scattering of two distinguishable scalars with the emission of a single gluon,
\begin{equation}
 \Phi_{1}(M,p_{1},a_{1}) + \phi_{2}(m,p_{2},a_{2}) \rightarrow \Phi_{3}(M,p_{3},a_{3}) + \phi_{4}(m,p_{4},a_{4}) + g(\epsilon,k,a).
\end{equation}
$\Phi$ and $\phi$ denotes the scalar particles and the $g$ denotes a gluon. The first label, for scalars, denote the masses of the particles and for the gluon it denotes the polarisation. The second label is for the $4$-momentum of the particle and last quantum number is for the color index of the particle. Here $\Phi_{1}$ and $\phi_{2}$ are taken to be incoming and $\Phi_{3}$ and $\phi_{4}$ are outgoing, with the gluon also outgoing.\\
The amplitude for the scattering process is
\begin{multline}
 \mathcal{A}(p_{3},p_{1},p_{4},p_{2},k) = g^{3}\bigg( \frac{C_{3}N_{3}}{t's'_{p_{3}k}}+\frac{C_{1}N_{1}}{t's'_{p_{1}k}}+\frac{C_{6}N_{6}}{t'} 
    +\frac{C_{4}N_{4}}{ts_{p_{4}k}}+\frac{C_{2}N_{2}}{ts_{p_{2}k}}+\frac{C_{5}N_{5}}{t}+\frac{C_{7}N_{7}}{t't}\bigg). \label{eq:22}
\end{multline}
The notation for the denominators are
\begin{align*}
 t  = (p_{1}-p_{3})^{2},\\
 t' = (p_{2}-p_{4})^{2},\\
 s'_{p_{1}k} = (p_{1}-k)^{2}-M^{2},\\
 s'_{p_{3}k} = (p_{3}+k)^{2}-M^{2},\\
 s_{p_{2}k} = (p_{2}-k)^{2}-m^{2},\\
 s_{p_{4}k} = (p_{4}+k)^{2}-m^{2}.
\end{align*}
The color factors are
\begin{align}
 C_{3} = f^{a_{3}ac}f^{cba_{1}}f^{a_{4}ba_{2}},\\
 C_{1} = f^{a_{3}bc}f^{caa_{1}}f^{a_{4}ba_{2}},\\
 C_{6} = f^{a_{3}ac}f^{cba_{1}}f^{a_{4}ba_{2}} + f^{a_{3}bc}f^{caa_{1}}f^{a_{4}ba_{2}},\\
 C_{4} = f^{a_{3}ba_{1}}f^{a_{4}ac}f^{cba_{2}},\\
 C_{2} = f^{a_{3}ba_{1}}f^{a_{4}bc}f^{caa_{2}},\\
 C_{5} = f^{a_{3}ba_{1}}f^{a_{4}ac}f^{cba_{2}} + f^{a_{3}ba_{1}}f^{a_{4}bc}f^{caa_{2}},\\
 C_{7} = f^{a_{3}ba_{1}}f^{bac}f^{a_{4}ca_{2}}.
\end{align}
These color factors satisfy the following Jacobi identities
\begin{align}
 C_{3} - C_{1} = C_{7}, \label{eq:20}\\
 C_{3} + C_{1} = C_{6}, \\
 C_{4} - C_{2} = -C_{7}, \label{eq:21}\\
 C_{4} + C_{2} = C_{5}.
\end{align}
The convention we have used for the adjoint representation of the Lie algebra generators is  $(T^{a}_{A})_{bc} = -if^{bac}$. 
Lastly the numerators are
\begin{flalign} \label{eq:28}
 N_{3} = 2i\big[\epsilon(k)\cdot p_{3}\big]\big[(p_{3}+k+p_{1})\cdot (p_{2}+p_{4})\big],\\
 N_{1} = 2i\big[\epsilon(k)\cdot p_{1}\big]\big[(p_{1}-k+p_{3})\cdot (p_{2}+p_{4})\big],\\
 N_{6} = -i(p_{2}+p_{4})\cdot \epsilon(k),\\
 N_{4} = 2i\big[\epsilon(k)\cdot p_{4}\big]\big[(p_{1}+p_{3})\cdot (p_{4}+k+p_{2})\big],\\
 N_{2} = 2i\big[\epsilon(k)\cdot p_{2}\big]\big[(p_{1}+p_{3})\cdot (p_{2}-k+p_{4})\big],\\
 N_{5} = -i(p_{1}+p_{3})\cdot \epsilon(k),\\
 N_{7} = \big[(p_{2}+p_{4})\cdot(p_{1}-p_{3}+k)\big]\big[(p_{1}+p_{3})\cdot\epsilon(k) \big] \label{eq:29} \\ 
         +\big[(p_{1}+p_{3})\cdot(p_{4}-p_{2}-k)\big]\big[(p_{2}+p_{4})\cdot\epsilon(k)\big] \\
         +\big[(p_{1}+p_{3})\cdot(p_{2}+p_{4})\big]\big[(p_{3}-p_{1}-p_{4}+p_{2})\cdot\epsilon(k)\big].
\end{flalign}
These numerators do not satisfy the color-kinematics duality relation. The first step to get it into the BCJ representation is to get rid of the diagrams with $4$-point vertices and the diagram with $3$-point Yang-Mills vertex. As in the pure YM case this is achieved by doing a generalised gauge transformation without changing the amplitude. The numerators after the gauge transformation are
\begin{align}
 N'_{3} = N_{3} + s'_{p_{3}k}\bigg(N_{6}+\frac{N_{7}}{2t}\bigg),\\
 N'_{1} = N_{1} + s'_{p_{1}k}\bigg(N_{6}-\frac{N_{7}}{2t}\bigg),\\
 N'_{6} = 0,\\
 N'_{4} = N_{4} + s_{p_{4}k}\bigg(N_{5}-\frac{N_{7}}{2t'}\bigg),\\
 N'_{2} = N_{2} + s_{p_{2}k}\bigg(N_{5}+\frac{N_{7}}{2t'}\bigg),\\
 N'_{5} = 0,\\
 N'_{7} = 0.
\end{align}
Inverting these transformations namely writing $N_{3}$ and $N_{1}$ in terms of $N'_{3},N'_{1},N_{6},N_{7}$ and $N_{4}$ and $N_{2}$ in terms of $N'_{4},N'_{2},N_{5},N_{7}$, plugging them into eq. \eqref{eq:22} and then using the Jacobi identities of the color factors we get
\begin{equation}
 \mathcal{A}(p_{3},p_{1},p_{4},p_{2},k) = g^{3}\bigg(\frac{C_{3}N'_{3}}{t's'_{p_{3}k}}+\frac{C_{1}N'_{1}}{t's'_{p_{1}k}}+\frac{C_{4}N'_{4}}{ts_{p_{4}k}}+\frac{C_{2}N'_{2}}{ts_{p_{2}k}}\bigg). \label{eq:24}
\end{equation}
The color factors now satisfy the identity
\begin{equation}
 C_{3}-C_{1}+C_{4}-C_{2}=0, \label{eq:23}
\end{equation}
which is obtained by eliminating $C_{6},C_{5}$ and $C_{7}$ using the Jacobi identities. The new numerators still do not satisfy the color-kinematics duality relation which now would be the kinematical analogue of eq.\eqref{eq:23}. To remedy this we do another generalised gauge transformation 
\begin{align}
 N''_{3} = N'_{3} + \alpha t's'_{p_{3}k},\\
 N''_{1} = N'_{1} - \alpha t's'_{p_{1}k},\\
 N''_{4} = N'_{4} + \alpha ts_{p_{4}k},\\
 N''_{2} = N'_{2} - \alpha ts_{p_{2}k}.
\end{align}
Here $\alpha$ is a function which determined by demanding that the new numerators satisfy the color-kinematics duality relation,
\begin{equation}
 N''_{3}-N''_{1}+N''_{4}-N''_{2}=0. \label{eq:25}
\end{equation}
So, we get 
\begin{equation} 
 \alpha = \frac{-N'_{3}+N'_{1}-N'_{4}+N'_{2}}{t'(s'_{p_{3}k}+s'_{p_{1}k})+t(s_{p_{4}k}+s_{p_{2}k})}.
\end{equation}
Finally plugging this back in to the eq.\eqref{eq:23}, we get
\begin{equation} \label{eq:26}
 \mathcal{A}(p_{3},p_{1},p_{4},p_{2},k) = g^{3}\bigg(\frac{C_{3}N''_{3}}{t's'_{p_{3}k}}+\frac{C_{1}N''_{1}}{t's'_{p_{1}k}}+\frac{C_{4}N''_{4}}{ts_{p_{4}k}}+\frac{C_{2}N''_{2}}{ts_{p_{2}k}}\bigg). 
\end{equation}
Now the amplitude is in a BCJ representation, namely all the numerators satisfy the color-kinematics duality relation of eq.\eqref{eq:25} and the corresponding identity for the color factors, eq.\eqref{eq:23}. The above result was derived in the case for massless scalar particles in \cite{vera}. Thus, we do a double copy ($C_{i}\rightarrow N''_{i}$) to get the corresponding gravitational amplitude,
\begin{equation} \label{eq:27}
 \mathcal{M}(p_{3},p_{1},p_{4},p_{2},k) = \bigg(\frac{\kappa}{2}\bigg)^{3}\bigg(\frac{N''_{3}N''_{3}}{t's'_{p_{3}k}}+\frac{N''_{1}N''_{1}}{t's'_{p_{1}k}}+\frac{N''_{4}N''_{4}}{ts_{q'k}}+\frac{N''_{2}N''_{2}}{ts_{p_{2}k}}\bigg).
\end{equation}
The above amplitude was shown to be the correct amplitude for scalars coupled to dilaton gravity in \cite{luna}.
We are interested in finding a double copy substitution to get the leading soft graviton theorem, so we will take the soft limit of eq.\eqref{eq:26}.
We note that the eq.\eqref{eq:26} is related to eq.\eqref{eq:24} by a gauge transformation hence, we will take the soft expansion of eq.\eqref{eq:24}. 
In doing this there are two contributions to the soft expansion, one is from the numerators ($N'_{i}$s) and the second from the propogators. The soft limit of the propogators are
\begin{align}
 s'_{p_{3}k} =\big[(p_{3}+k)^{2}-M^{2}\big] =  \lim_{\omega \rightarrow 0} \omega(2p_{3}\cdot\hat{k}) \rightarrow 0, \\
 s'_{p_{1}k} =\big[(p_{1}-k)^{2}-M^{2}\big] =  \lim_{\omega \rightarrow 0} \omega(-2p_{1}\cdot\hat{k}) \rightarrow 0, \\
 s'_{p_{4}k} =\big[(p_{4}+k)^{2}-m^{2}\big] =  \lim_{\omega \rightarrow 0} \omega(2p_{4}\cdot\hat{k}) \rightarrow 0, \\
 s'_{p_{2}k} =\big[(p_{2}-k)^{2}-m^{2}\big] =  \lim_{\omega \rightarrow 0} \omega(-2p_{2}\cdot\hat{k}) \rightarrow 0.
\end{align}
The contribution from the gluon mediating the interaction between the scalar particles doesn't contribute to the leading soft gluon theorem, hence, $t=t'$. For the soft expansion of the numerators we will keep only the terms which are $\mathcal{O}(1)$ as these are the ones which contribute to the leading term ($\mathcal{O}(\frac{1}{\omega})$) in the amplitude, hence,
\begin{align}
 N'_{3} = \lim_{\omega \rightarrow 0} \bigg[ N_{3} + \omega(2p_{3}\cdot\hat{k})\bigg(N_{6} + \frac{N_{7}}{2t'}\bigg) \bigg] = N_{3} + \mathcal{O}(\omega), \\
 N'_{1} = \lim_{\omega \rightarrow 0} \bigg[ N_{1} - \omega(2p_{1}\cdot\hat{k})\bigg(N_{6} - \frac{N_{7}}{2t'}\bigg) \bigg] = N_{1} + \mathcal{O}(\omega),\\
 N'_{4} = \lim_{\omega \rightarrow 0} \bigg[ N_{4} + \omega(2p_{4}\cdot\hat{k})\bigg(N_{5} - \frac{N_{7}}{2t'}\bigg) \bigg] = N_{4} + \mathcal{O}(\omega),\\
 N'_{2} = \lim_{\omega \rightarrow 0} \bigg[ N_{2} -\omega(2p_{2}\cdot\hat{k})\bigg(N_{5} - \frac{N_{7}}{2t'}\bigg) \bigg] = N_{2} + \mathcal{O}(\omega).
\end{align}
The terms multiplying the propogators don't contain any terms which are of $\mathcal{O}(\frac{1}{\omega})$ as can be seen by examining the expressions for $N_{6},N_{5}$ and $N_{7}$ from the eq.\eqref{eq:28}-\eqref{eq:29}.\\
Hence, under the soft expansion the amplitude in eq.\eqref{eq:24} becomes
\begin{equation}
 \mathcal{A}(p_{3},p_{1},p_{4},p_{2},k) = g^{3}\bigg(\frac{C_{3}N_{3}}{t's'_{p_{3}k}}+\frac{C_{1}N_{1}}{t's'_{p_{1}k}}+\frac{C_{4}N_{4}}{t's_{p_{4}k}}+\frac{C_{2}N'_{2}}{t's_{p_{2}k}}\bigg) +\mathcal{O}(1).
\end{equation}
The $N_{i}$s are
\begin{align}
 N_{3} = (2i)\mathcal{A}(p_{3},p_{1},p_{4},p_{2}) \epsilon(k)\cdot p_{3}+\mathcal{O}(\omega), \\
 N_{1} = (2i)\mathcal{A}(p_{3},p_{1},p_{4},p_{2}) \epsilon(k)\cdot p_{1} +\mathcal{O}(\omega),\\
 N_{4} = (2i)\mathcal{A}(p_{3},p_{1},p_{4},p_{2}) \epsilon(k)\cdot p_{4} +\mathcal{O}(\omega),\\
 N_{2} = (2i)\mathcal{A}(p_{3},p_{1},p_{4},p_{2}) \epsilon(k)\cdot p_{2} +\mathcal{O}(\omega),\\
 \mathcal{A}(p_{3},p_{1},p_{4},p_{2}) = (p_{1}+p_{3})\cdot (p_{2}+p_{4}). \label{nuhard}
\end{align}
Plugging this into the previous equation and expanding the denominator,
\begin{multline}
 \mathcal{A}(p_{3},p_{1},p_{4},p_{2},k) =\\
 \frac{g}{\omega}\bigg[C_{3}\frac{\epsilon(k)\cdot p_{3}}{(p_{3}\cdot\hat{k})} + C_{1}\frac{\epsilon(k)\cdot p_{1}}{(-p_{1}\cdot\hat{k})}  
 + C_{4}\frac{\epsilon(k)\cdot p_{4}}{(p_{4}\cdot\hat{k})} + C_{2}\frac{\epsilon(k)\cdot p_{2}}{(-p_{2}\cdot\hat{k})}\bigg]\frac{(ig^{2})\mathcal{A}(p_{3},p_{1},p_{4},p_{2})}{t'} + \mathcal{O}(1).
\end{multline}
Hence,
\begin{multline} \label{eq:139}
 \lim_{\omega \rightarrow 0} \omega\mathcal{A}(p_{3},p_{1},p_{4},p_{2},k)
  =\\
  g\bigg[C_{3}\frac{\epsilon(k)\cdot p_{3}}{(p_{3}\cdot\hat{k})} + C_{1}\frac{\epsilon(k)\cdot p_{1}}{(-p_{1}\cdot \hat{k})} 
 + C_{4}\frac{\epsilon(k)\cdot p_{4}}{(p_{4}\cdot\hat{k})} + C_{2}\frac{\epsilon(k)\cdot p_{2}}{(-p_{2}\cdot\hat{k})}\bigg]\frac{(ig^{2})\mathcal{A}(p_{3},p_{1},p_{4},p_{2})}{t'}.
\end{multline}
The last equation can also be written as
\begin{multline} \label{eq:36}
 \lim_{\omega \rightarrow 0} \omega\mathcal{A}^{a_{3},a_{1},a_{4},a_{2},a}(p_{3},p_{1},p_{4},p_{2},k)
  = \\g\bigg[(T^{a}_{3})_{A}\frac{\epsilon\cdot p_{3}}{(p_{3}\cdot\hat{k})}+(T^{a}_{1})_{A}\frac{\epsilon\cdot p_{1}}{(-p_{1}\cdot \hat{k})} 
    + (T^{a}_{4})_{A}\frac{\epsilon\cdot p_{4}}{(p_{4}\cdot\hat{k})} + (T^{a}_{2})_{A}\frac{\epsilon\cdot p_{2}}{(-p_{2}\cdot\hat{k})}\bigg]\mathcal{A}^{a_{3},a_{1},a_{4},a_{2}}(p_{3},p_{1},p_{4},p_{2}).
\end{multline}
This is the leading soft gluon theorem for two incoming and two outgoing particles. The amplitude $\mathcal{A}^{a_{3},a_{1},a_{4},a_{2}}(p_{3},p_{1},p_{4},p_{2})$ is the $4$-point amplitude without the soft gluon and the $(T^{a}_{i})_{A}$ is the Lie algebra generator, in the adjoint representation, acting on the $i^{th}$ particle in the $4$-point amplitude.\\
The color factors in eq.\eqref{eq:139} satisfy the identity in eq.\eqref{eq:23}. The corresponding numerator relation is  
\begin{equation}\label{leading}
 \epsilon(k)\cdot (p_{3}-p_{1}+p_{4}-p_{2})\ (ig^{2})\mathcal{A}(p_{3},p_{1},p_{4},p_{2}) = 0.
\end{equation}
In the above equation the $\mathcal{A}(p_{3},p_{1},p_{4},p_{2})$ is the numerator of the hard amplitude which is given in eq.\eqref{nuhard}, which is common to all the terms in eq.\eqref{eq:139}. This means that the color-kinematics duality is satisfied due to momentum conservation of the terms in the numerators of the leading soft factor,
\begin{equation}
 \epsilon(k)\cdot(p_{3}-p_{1}+p_{4}-p_{2}) = 0.
\end{equation}
Hence we see that it is not only the whole amplitude, but the Weinberg soft factor by itself has the manifest color-kinematics duality and hence we can apply the (quantum) double copy to the soft factor itself!\\
This clearly yields the leading soft graviton factor,
\begin{equation} \label{eq:35}
 \lim_{\omega \rightarrow 0} \omega\mathcal{M}_{5}(p_{3},p_{1},p_{4},p_{2},k) = \bigg(\frac{\kappa}{2}\bigg)^{3}\bigg[\frac{(\epsilon\cdot p_{3})^{2}}{p_{3}\cdot\hat{k}}-\frac{(\epsilon\cdot p_{1})^{2}}{p_{1}\cdot\hat{k}}+\frac{(\epsilon\cdot p_{4})^{2}}{p_{4}\cdot\hat{k}}-\frac{(\epsilon\cdot p_{2})^{2}}{p_{2}\cdot\hat{k}}\bigg]\mathcal{M}_{4}(p_{3},p_{1},p_{4},p_{2}).
\end{equation}
Here the $4$-point amplitude 
\begin{equation}
 \mathcal{M}_{4}(p_{3},p_{1},p_{4},p_{2}) = \frac{-\mathcal{A}^{2}(p_{3},p_{1},p_{4},p_{2})}{t'}.
\end{equation}
\begin{itemize}
 \item  Based on the extended studies of the quantum double copy, we know that the resulting $4$-point amplitude, $\mathcal{M}_{4}(p_{3},p_{1},p_{4},p_{2})$,   
will also contain a dilaton field in this case (when matter particles are scalars) and an axion field more generally. However, this caveat is not important when we consider the low frequency classical radiation which is independent of the details of the scattering and only depends on soft factors.
\item The universality of the single leading soft gluon theorem in $D\  \textgreater \ 4$ ensures that it maps to the single soft graviton factor, which means that we have effectively isolated the color-kinematics substitution for the leading soft graviton factor.
\end{itemize}
\begin{equation}
\label{eq:3}
 T^{a}_{i}\rightarrow p_{i}^{\mu}.
\end{equation}
The substitution is to replace the Lie algebra generator acting on the $i^{th}$ particle's color index with the same particle's $4$-momentum. We call this replacement rule, the \textit{soft double copy}, which proves one of the two classical double copy relations in the soft limit.\\ 
For classical colored objects which are in a color coherent state, this implies that 
\begin{equation}
 c^{a}_{i}\rightarrow  p_{i}^{\mu}.
\end{equation}
One can now use this substitution to relate the radiative gauge field in eq.\eqref{eq:120} to the far field metric in eq.\eqref{eq:121} which was obtained by taking the low frequency limit of the result \cite{goldberger}. Hence, as a result we have derived the one of the classical double copy substitution conjectured by \cite{goldberger}.


\subsection{Subleading soft gluon theorem}\label{5a}

In this section we will show that the soft double copy we derived for the leading soft gluon factor holds for the subleading factor as well.\\
The strategy is the same as in the leading case, namely, we start with the 5-point amplitude in the BCJ representation given in eq.\eqref{eq:26}, do the soft expansion upto the subleading order and check if the soft double copy derived for the leading soft factor holds at the subleading order as well.\\ 
To write the amplitude eq.\eqref{eq:26} in terms of soft factors till the subleading order we will use a well known result, namely, the subleading soft gluon factor for tree level amplitudes in Yang-Mills and scalar QCD is completely fixed by gauge invariance \cite{vera1,bern2,broedel}.\\ 
Further it was shown in \cite{vera1} that for tree level scattering of scalars coupled to Yang-Mills the single soft subleading factors for the gluon and the graviton can be written in such a form that the double copy structure is manifest between the two, but a color-kinematics duality for the amplitude was lacking.\\
Hence, following the analysis done in\cite{vera1}, the amplitude in eq.\eqref{eq:26} can be written as
\begin{multline} \label{eq:107}
 \mathcal{A}(p_{3},p_{1},p_{4},p_{2},k) =\\ 
 g\bigg[C_{1}\frac{\epsilon\cdot p_{1}}{(-p_{1}\cdot k)} + C_{2}\frac{\epsilon\cdot p_{2}}{(-p_{2}\cdot k)} + C_{3}\frac{\epsilon\cdot p_{3}}{(p_{3}\cdot k)} +  
 + C_{4}\frac{\epsilon\cdot p_{4}}{(p_{4}\cdot k)} - C_{1}\frac{\epsilon_{\mu}k_{\nu}J^{\mu\nu}_{(1)}}{(-p_{1}\cdot k)} \\
 - C_{2}\frac{\epsilon_{\mu}k_{\nu}J^{\mu\nu}_{(2)}}{(-p_{2}\cdot k)} + C_{3}\frac{\epsilon_{\mu}k_{\nu}J^{\mu\nu}_{(3)}}{(p_{3}\cdot k)} + C_{4}\frac{\epsilon_{\mu}k_{\nu}J^{\mu\nu}_{(4)}}{(p_{4}\cdot k)}\bigg]\frac{(ig^{2})\mathcal{A}(p_{3},p_{1},p_{4},p_{2})}{t'} + \mathcal{O}(\omega).
\end{multline}
where the angular momentum operator is given in eq.\eqref{eq:111}. The soft gluon theorem at subleading order in soft momentum is 
\begin{multline}\label{subleading1}
 \mathcal{A}^{a_{3},a_{1},a_{4},a_{2},a}(p_{3},p_{1},p_{4},p_{2},k) = \\
 g\bigg[(T_{1}^{a})_{A}\frac{\epsilon\cdot p_{1}}{(-p_{1}\cdot k)} + (T_{2}^{a})_{A}\frac{\epsilon\cdot p_{2}}{(-p_{2}\cdot k)} + (T_{3}^{a})_{A}\frac{\epsilon\cdot p_{3}}{(p_{3}\cdot k)} +  
 + (T_{4}^{a})_{A}\frac{\epsilon\cdot p_{4}}{(p_{4}\cdot k)}- (T_{1}^{a})_{A}\frac{\epsilon_{\mu}k_{\nu}J^{\mu\nu}_{(1)}}{(-p_{1}\cdot k)} \\
  - (T_{2}^{a})_{A}\frac{\epsilon_{\mu}k_{\nu}J^{\mu\nu}_{(2)}}{(-p_{2}\cdot k)} + (T_{3}^{a})_{A}\frac{\epsilon_{\mu}k_{\nu}J^{\mu\nu}_{(3)}}{p_{3}\cdot k} + (T_{4}^{a})_{A}\frac{\epsilon_{\mu}k_{\nu}J^{\mu\nu}_{(4)}}{(p_{4}\cdot k)}\bigg]\mathcal{A}^{a_{3},a_{1},a_{4},a_{2}}(p_{3},p_{1},p_{4},p_{2}) + \mathcal{O}(\omega).
\end{multline}
From the eq.\eqref{eq:107} the numerator relation from the color-kinematics relation \eqref{eq:25} at the subleading order is 
\begin{equation}\label{eq:106}
 \sum_{i=1}^{4} J_{(i)}^{\mu\nu}\mathcal{A}(p_{3},p_{1},p_{4},p_{2}) = 0.
\end{equation}
To check this we use the following relations 
\begin{align}
 J^{\mu\nu}_{(1)}\ \mathcal{A}(p_{3},p_{1},p_{4},p_{2}) =  \frac{p_{1}^{\mu}(p_{2}+p_{4})^{\nu} - p_{1}^{\nu}(p_{2}+p_{4})^{\mu}}{t'},\\
 J^{\mu\nu}_{(2)}\ \mathcal{A}(p_{3},p_{1},p_{4},p_{2}) = \frac{p_{2}^{\mu}(p_{1}+p_{3})^{\nu} - p_{2}^{\nu}(p_{1}+p_{3})^{\mu}}{t},\\
 J^{\mu\nu}_{(3)}\ \mathcal{A}(p_{3},p_{1},p_{4},p_{2}) = \frac{p_{3}^{\mu}(p_{2}+p_{4})^{\nu} - p_{3}^{\nu}(p_{2}+p_{4})^{\mu}}{t'},\\
 J^{\mu\nu}_{(4)}\ \mathcal{A}(p_{3},p_{1},p_{4},p_{2}) = \frac{p_{4}^{\mu}(p_{1}+p_{3})^{\nu} - p_{4}^{\nu}(p_{1}+p_{3})^{\mu}}{t}. 
\end{align}
Here we have used momentum conservation for the hard amplitude to write $t=t'$. Summing all of these terms, the L.H.S of the eq.\eqref{eq:106} becomes
\begin{multline}
 = \frac{1}{t} \big[ (p_{1}+p_{3})^{\mu}(p_{2}+p_{4})^{\nu} - (p_{1}+p_{3})^{\nu}(p_{2}+p_{4})^{\mu}\\ 
  + (p_{2}+p_{4})^{\mu}(p_{1}+p_{3})^{\nu} - (p_{2}+p_{4})^{\nu}(p_{1}+p_{3})^{\mu} \big] = 0.   
\end{multline}
What we have verified above is that the color-kinematics duality for the full amplitude,\\$\mathcal{A}^{a_{1},a_{2},a_{3},a_{4},a}(p_{1},p_{2},p_{3},p_{4},k)$, in the soft limit is preserved at the subleading order in soft momentum.\\
At this stage in the previous section, we saw below eq.\eqref{leading} that the Weinberg soft factor itself satisfies the duality due to momentum conservation. We now see that due to angular momentum conservation, even the subleading factor has this symmetry.
More in detail, due to  
\begin{equation}
\sum_{i=1}^{4} \epsilon_{\mu}k_{\nu}J_{(i)}^{\mu\nu}\ =\ 0,
\end{equation}
it can be readily verified that subleading soft gluon factor in eq.\eqref{subleading1} also has a manifest color-kinematics duality symmetry. 
This in turn proves the classical double copy relations between gluon and gravitational radiation upto subleading order in radiation frequency.
\begin{multline}
\mathcal{M}_{5}(p_{1},p_{2},p_{3},p_{4},k) = 
\bigg(\frac{\kappa}{2}\bigg)^{3}\bigg[-\frac{(\epsilon\cdot p_{1})^{2}}{p_{1}\cdot k}  -\frac{(\epsilon\cdot p_{2})^{2}}{p_{2}\cdot k} + \frac{(\epsilon\cdot p_{3})^{2}}{(p_{3}\cdot k)} +  
\frac{(\epsilon\cdot p_{4})^{2}}{(p_{4}\cdot k)} \\
+ \frac{\epsilon_{\mu\nu}p_{1}^{\mu}k_{\rho}J^{\nu\rho}_{(1)}}{p_{1}\cdot k} + \frac{\epsilon_{\mu\nu}p_{2}^{\mu}k_{\rho}J^{\nu\rho}_{(2)}}{p_{2}\cdot k} + \frac{\epsilon_{\mu\nu}p_{3}^{\mu}k_{\rho}J^{\nu\rho}_{(3)}}{p_{3}\cdot k} + \frac{\epsilon_{\mu\nu}p_{4}^{\mu}k_{\rho}J^{\nu\rho}_{(4)}}{p_{4}\cdot k}\bigg]\mathcal{M}_{4}(p_{1},p_{2},p_{3},p_{4}) + \mathcal{O}(\omega).
\end{multline}
As we saw in the previous section for classical colored particles the gluon generators, $T_{i}^{a}$, are replaced with the color charges, $c_{i}^{a}$. Doing this substitution in eq.\eqref{subleading1}, we see that the soft double copy does produce the correct soft graviton factor at the subleading order in soft momemtum.\\
Thus we have finally arrived at our main result : The classical double copy which maps the the radiative gauge field in eq.\eqref{subgauge} to the far field metric in eq.\eqref{subgrav} can be derived from color-kinematics duality upto subleading order in the frequency of radiation.


\section{Conclusion and Outlook}\label{6}

Classical gluon and gravitational radiation are phenomenologically diverce phenomena in nature. Whereas the gluon radiation is relevant in ultra-relativistic nuclei collisions \cite{kovchegov}, observable classical gravitational radiation is produced in astrophysical processes like black-hole collisions! However inspite of the mutually exclusive domains of their existence, they seem to be rather closely intertwined in the theoretical domain via the classical double copy.\\
In this paper we attempted to derive one of the two classical double copy relations, namely $c^{a}\ \rightarrow\ p^{\mu}$ from the color-kinematics duality of scattering amplitudes. We showed that the low frequency gravitational radiation upto subleading order is indeed a double copy of the soft gluon radiation and that this double copy is simply a manifestation of the color-kinematics duality of the scattering amplitudes. The key ideas used in our analysis are already well understood in the literature. They are ;\noindent{ (1)} deriving soft radiation from soft theorems for gravity, \noindent{ (2)} modelling classical color charges using the McLerran-Venugopalan model, and \noindent{ (3)} color-kinematics duality for tree-level amplitudes in scalar QCD. All three of these come together to conspire a rather simple proof of the classical double copy at the subleading order in the soft limit.\\
Coherent state of asymptotic color charges implied that the multiple soft gluon theorem was abelianised" upto subleading order and hence, the analysis of \cite{ashoke} was directly applicable to obtain the radiative gluon field (at subleading order in radiation frequency) from soft theorems.  An immediate consequence of this was a direct relationship between low frequency radiation and single soft factor not only for gravity but also for gluon radiation. Whence the classical double copy rule was reduced to showing the double copy of single soft factors. Although our analysis is rather different and restricted to a simpler setting of low frequency radiation, we believe this result is consistent with proof of the classical double copy obtained in \cite{guevara}.\\
We then proceeded to see if a first principle derivation of the double copy construction of soft graviton factor from soft gluon factor could be given using color-kinematics duality. Writing the (tree-level) scalar QCD amplitude in BCJ representation led us to a nice result that the leading soft gluon factor also had a manifest color-kinematics duality symmetry. This proves the classical double copy relation for the leading order soft radiation from BCJ representation of quantum amplitudes! We note that although the color-kinematics duality is only proved for scalar QCD amplitudes at tree level, as the leading soft factor is universal in $D\  \textgreater \ 4$ dimensions, our proof will remain valid even when loop corrections are taken into account.\\
As we saw in section \ref{5a}, proof of the classical double copy from color-kinematics duality also remained valid at subleading order in the large impact parameter regime. For the class of scattering processes such as the probe-scatterer limit in \cite{ashoke},  the non-universal terms in the subleading soft gluon factor contribute at the same order as the universal term. As the subleading soft graviton factor is universal in $D\ >\ 4$ dimensions, it remains unclear how our proof of classical double copy can be extended to this case. We leave this question for future investigations. It is plausible that the second classical double copy rule which maps structure constant of the gauge group to three point vertex in  Yang-Mills theory may feature in ``douple copying" the non-universal terms. This is because the non-universal terms at the subleading order depend on the internal structure of the scatterer which in turn involves just such three point couplings. As the subleading soft gravitational radiation is universal, a ``double copy" rule which maps the gluon radiation in such processes to soft gravitational radiation may shed light on the full set of classical double copy relations.\\
Many other questions remain open. The recent work of \cite{guevara} proves the classical double copy from scattering amplitudes rather beautifully using the so called ``soft exponentiation" of scattering amplitudes. It will be extremely interesting to consider the relationship between the two approaches of proving classical double copy from quantum amplitudes. Our analysis for the leading soft gluon factor was valid for $D\  \textgreater \ 4$ dimensions. Although the leading soft graviton factor does not change even in $D\ =\ 4$, the leading gluon factor does receive infra-red divergent loop corrections. It will be interesting to revisit the classical double copy for radiation in $D\ =\ 4$ and check if indeed it can be proved using color-kinematics duality.

\section{Ackowledgements}

We would like to thank Alok Laddha for proposing the problem, many useful discussions and helping with the manuscript. We are thankful to Pinaki Banerjee for carefully going through the manuscript and suggesting several changes. We would also like to thank IISER Bhopal and ST4 (Student Talks on Trending Topics in theory) for the hospitality provided during the completion of the manuscript. AM would also like to thank Sudipta Sarkar and IIT Gandhinagar for their generous hospitality during the course of the project. We would also like to thank Alexander Ochirov for many useful comments on the earlier manuscript.    

\appendix 
\section{Classical soft gluon factor at subleading order}\label{slglu}

In this appendix we show that the subleading expansion of gluon field given in eqns. (\ref{eq:45}, \ref{mudef}) is proportional to classical subleading soft gluon factor. \\
We start with the known definition of the subleading soft factor. Given a  particle $i$ with initial/final position, momenta and color charges given by $x^{\pm \mu}_{i},\ p^{\pm \mu}_{i}$ and $c^{\pm a}_{i}$ , the soft factor is given by,
\begin{equation}
S^{(1) \mu}_{a}\ =\ \sum_{i}\ c^{+ a}_{i}\ \frac{\bold{J}^{+ \mu\nu}_{i}\ k_{\nu}}{p_{i}^{+}\cdot k}\ -\ c_{i}^{- a}\ \frac{\bold{J}_{i}^{- \mu\nu}k_{\nu}}{p_{i}^{-}\cdot k}.
\end{equation}
Upto leading order in the coupling $g$, we can simplify this expression as, 
\begin{align}\label{g2ssoft}
S^{(1) \mu}_{a} =
\sum_{i}\frac{c_{i}^{- a}}{p_{i}^{-}\cdot k}\left[ (\bold{J}_{i}^{+ \mu\nu} - \bold{J}_{i}^{- \mu\nu})\cdot k_{\nu} - \bold{J}_{i}^{- \mu\nu}k_{v} \left(\frac{\dot{z}^{+}\cdot k}{v_{i}^{-}\cdot k}\right) \right]
+\ \sum_{i} (c_{i}^{+ a} - c_{i}^{- a}) \frac{\bold{J}_{i}^{- \mu\nu}k_{\nu}}{p_{i}^{-}\cdot k}.
\end{align}
We will now show that the first term in the eqn.(\ref{g2ssoft}) equals the sum of all the subleading terms in eqn.(\ref{eq:45}) barring the term containing $f^{abc}$. Similarly the second term equals the subleading contribution of the term proportional to $f^{abc}$. \\
Using $x^{\mu}_{i}(\tau)\ =\ b_{i}^{- \mu}\ +\ v^{- \mu}\tau\ +\ z^{\mu}(\tau)$ with $z^{\mu}(\tau^{-})\ =\ 0$, it can be easily shown that 
\begin{equation}\label{diffj}
\bold{J}_{i}^{+ \mu\nu}\ -\ \bold{J}_{i}^{- \mu\nu}\ =\ b_{i}^{- \mu}\dot{z}_{i}^{+ \nu}\ -\ b_{i}^{- \nu}\dot{z}_{i}^{+ \mu},
\end{equation}
where $\dot{z}^{+ \mu}\ :=\ \dot{z}^{\mu}(\tau^{+})$. Using eqn.(\ref{diffj}) and after some algebra, the first term of eqn.(\ref{g2ssoft}) simplifies and yields,
\begin{equation}
S^{(1) \mu}_{a}\ =
\sum_{i}\ c_{i}^{- a}\ \left(\frac{b^{-}_{i}\cdot k}{v_{i}^{-}\cdot k}\right)\ \left[\ \left(\frac{\dot{z}^{+}_{i}\cdot k}{v_{i}^{-}\cdot k}\right)v_{i}^{- \mu}\ -\ \dot{z}_{+}^{\mu}\ \right]
+\ \sum_{i} (c_{i}^{+ a}\ -\ c_{i}^{- a})\ \frac{\bold{J}_{i}^{-\ \mu\nu}k_{\nu}}{p_{i}^{-}\cdot k}.
\end{equation}
Using the equation of motion for the $i^{th}$ particle, and using the fact that $\dot{z}^{\mu}(\tau^{-})\ =\ 0$, we get,
\begin{equation}
\dot{z}_{i}^{+ \mu}\ =\ \frac{-ig^{2}}{m_{i}}\sum_{j\neq i}\int_{l_{j}}\ (c^{-}_{i}\cdot c^{-}_{j})\ (2\pi)^{2}\ \frac{\delta(l_{j}\cdot v_{j}^{-})\ \delta(l_{j}\cdot v_{i}^{-})}{l_{j}^{2}}\ [\ (v_{i}^{-}\cdot v_{j}^{-})\ l_{j}^{\mu}\ -\ (v_{i}^{-}\cdot l_{j})\ v_{j}^{- \mu}\ ],
\end{equation}
substituting $\dot{z}_{i}^{+ \mu}$ in the previous equation we get the following integral form for the ``abelian" term in the soft factor as, 
\begin{equation}\label{clusf}
\begin{array}{lll}
S^{(1) \mu}_{a}\ =\\ 
\sum_{i}\ \frac{-ig^{2}}{m_{i}}\ c_{i}^{- a} \left(\frac{b^{-}_{i}\cdot k}{v_{i}^{-}\cdot k}\right)
 \sum_{j\neq i}\int_{l_{j}}\ (c^{-}_{i}\cdot c^{-}_{j})\ (2\pi)^{2}\ \frac{\delta(l_{j}\cdot v_{j}^{-})\ \delta(l_{j}\cdot v_{i}^{-})}{l_{j}^{2}}\ \Big[\ v_{i}^{- \mu}\{\ \frac{v_{i}^{-}\cdot v_{j}^{-}}{v_{i}^{-}\cdot k}\ l_{j}\cdot k\ -\ \frac{(v_{i}\cdot l_{j})(v_{j}\cdot l_{j})}{v_{i}\cdot k}\ \}.\\
.\hspace*{4.2in} -\ \{ (v_{i}^{-}\cdot v_{j}^{-}) l_{j}^{\mu}\ -\ (v_{i}^{-}\cdot l_{j})\ v_{j}^{\mu}\ \}\Big] \\
\hspace*{2.4in}+\ \sum_{i} (c_{i}^{+ a}\ -\ c_{i}^{- a})\ \frac{\bold{J}_{i}^{-\ \mu\nu}k_{\nu}}{p_{i}^{-}\cdot k}.
\end{array}
\end{equation}
By expanding eqn.(\ref{mudef}) to subleading order in $k$ we can compare the first term of eqn.(\ref{clusf}) with the ``abelian" term (that does not contain $f^{abc}$) in eqn.(\ref{eq:45}).\\
Subleading expansion of $\mu_{i,j}(k)$ in eqn.(\ref{mudef}) is given by, 
\begin{equation}\label{mudefsub}
\mu_{i,j}(k)\ =\ i\ (2\pi)^{D}\ (2\pi)^{2}\ \frac{\delta(v_{i}^{-}\cdot l_{i})\delta(v_{j}^{-}\cdot l_{j})}{l_{i}^{2}l_{j}^{2}} (b_{-}^{i}\cdot k)\ \left[\frac{1}{v_{i}^{-}\cdot k}\ +\ 1\right]\ \delta^{D}(l_{i}\ +\ l_{j})
\end{equation}
The leading term in the above equation comes from the integration region of $l_{i} \sim k$ and the subleading contribution comes from finite $l_{i}$ region. \\
It is now easy to check that, substituting eqn.(\ref{mudefsub}) in the first term of eqn.(\ref{eq:45}) precisely equals the first term of eqn.(\ref{clusf}). Matching of the non-abelian term in eqn.(\ref{eq:45}) with eqn.(\ref{clusf}) proceeds in an exactly analogous manner.\\ 
Thus the radiative gluon field at subleading order matches with the classical soft factor.


\newpage

\begin{thebibliography}{99}

\bibitem{cachazo} 
  F.~Cachazo, S.~He and E.~Y.~Yuan,
  ``Scattering equations and Kawai-Lewellen-Tye orthogonality,''
  Phys.\ Rev.\ D {\bf 90}, no. 6, 065001 (2014)
  doi:10.1103/PhysRevD.90.065001
  \href{https://arxiv.org/abs/1306.6575}{{\tt arXiv:1306.6575 [hep-th]}}.
  
\bibitem{bern} 
  Z.~Bern, J.~J.~M.~Carrasco and H.~Johansson,
  ``New Relations for Gauge-Theory Amplitudes,''
  Phys.\ Rev.\ D {\bf 78}, 085011 (2008)
  doi:10.1103/PhysRevD.78.085011
   \href{https://arxiv.org/abs/0805.3993}{{\tt arXiv:0805.3993 [hep-th]}}.
  
\bibitem{goldberger} 
  W.~D.~Goldberger and A.~K.~Ridgway,
  ``Radiation and the classical double copy for color charges,''
  Phys.\ Rev.\ D {\bf 95}, no. 12, 125010 (2017)
  doi:10.1103/PhysRevD.95.125010
  \href{https://arxiv.org/abs/1611.03493}{{\tt arXiv:1611.03493 [hep-th]}}.

\bibitem{ridgway} 
  W.~D.~Goldberger and A.~K.~Ridgway,
  ``Bound states and the classical double copy,''
  Phys.\ Rev.\ D {\bf 97}, no. 8, 085019 (2018)
  doi:10.1103/PhysRevD.97.085019
  \href{https://arxiv.org/abs/1711.09493}{{\tt arXiv:1711.09493 [hep-th]}}.

\bibitem{plefka2} 
  J.~Plefka, J.~Steinhoff and W.~Wormsbecher,
  ``Effective action of dilaton gravity as the classical double copy of Yang-Mills theory,''
  Phys.\ Rev.\ D {\bf 99}, no. 2, 024021 (2019)
  doi:10.1103/PhysRevD.99.024021
  \href{https://arxiv.org/abs/1807.09859}{{\tt arXiv:1807.09859 [hep-th]}}.

\bibitem{monteiro} 
  R.~Monteiro, D.~O'Connell and C.~D.~White,
  ``Black holes and the double copy,''
  JHEP {\bf 1412}, 056 (2014)
  doi:10.1007/JHEP12(2014)056
  \href{https://arxiv.org/abs/1410.0239}{{\tt arXiv:1410.0239 [hep-th]}}.

\bibitem{white1} 
  A.~Luna, R.~Monteiro, D.~O'Connell and C.~D.~White,
  ``The classical double copy for Taub–NUT spacetime,''
  Phys.\ Lett.\ B {\bf 750}, 272 (2015)
  doi:10.1016/j.physletb.2015.09.021
\href{https://arxiv.org/abs/1507.01869}{{\tt arXiv:1507.01869 [hep-th]}}.  
  
\bibitem{white2} 
  A.~Luna, R.~Monteiro, I.~Nicholson, D.~O'Connell and C.~D.~White,
  ``The double copy: Bremsstrahlung and accelerating black holes,''
  JHEP {\bf 1606}, 023 (2016)
  doi:10.1007/JHEP06(2016)023
\href{https://arxiv.org/abs/1603.05737}{{\tt arXiv:1603.05737 [hep-th]}}.  
  
\bibitem{nicholson} 
A.~Luna, R.~Monteiro, I.~Nicholson, A.~Ochirov, D.~O'Connell, N.~Westerberg and C.~D.~White,
``Perturbative spacetimes from Yang-Mills theory,''
JHEP {\bf 1704}, 069 (2017)
doi:10.1007/JHEP04(2017)069
\href{https://arxiv.org/abs/1611.07508}{{\tt arXiv:1611.07508 [hep-th]}}
  
\bibitem{white3} 
  N.~Bahjat-Abbas, A.~Luna and C.~D.~White,
  ``The Kerr-Schild double copy in curved spacetime,''
  JHEP {\bf 1712}, 004 (2017)
  doi:10.1007/JHEP12(2017)004
\href{https://arxiv.org/abs/1710.01953}{{\tt arXiv:1710.01953 [hep-th]}}.  
  
\bibitem{white4} 
  D.~S.~Berman, E.~Chacón, A.~Luna and C.~D.~White,
  ``The self-dual classical double copy, and the Eguchi-Hanson instanton,''
  JHEP {\bf 1901}, 107 (2019)
  doi:10.1007/JHEP01(2019)107
\href{https://arxiv.org/abs/1809.04063}{{\tt arXiv:1809.04063 [hep-th]}}.  
  
\bibitem{white5} 
  M.~Carrillo González, B.~Melcher, K.~Ratliff, S.~Watson and C.~D.~White,
  ``The classical double copy in three spacetime dimensions,''
\href{https://arxiv.org/abs/1904.11001}{{\tt arXiv:1904.11001 [hep-th]}}.
  
  
\bibitem{plefka1}
 J.~Plefka and C.~Shi, J.~Steinhoff and T.~Wang
 ``Breakdown of the classical double copy for the effectiv action of dilaton-gravity at NNLO"
\href{}{{\tt arXiv:1906.05875}} 
 
\bibitem{shen}
  C.~H.~Shen,
  ``Gravitational Radiation from Color-Kinematics Duality,''
  JHEP {\bf 1811}, 162 (2018)
  doi:10.1007/JHEP11(2018)162
  \href{https://arxiv.org/abs/1806.07388}{{\tt arXiv:1806.07388 [hep-th]}}.
  
  
 \bibitem{guevara} 
  Y.~F.~Bautista and A.~Guevara,
  ``From Scattering Amplitudes to Classical Physics: Universality, Double Copy and Soft Theorems,''
  \href{https://arxiv.org/abs/1903.12419}{{\tt arXiv:1903.12419 [hep-th]}}.
  
\bibitem{luna} 
  A.~Luna, I.~Nicholson, D.~O'Connell and C.~D.~White,
  ``Inelastic Black Hole Scattering from Charged Scalar Amplitudes,''
  JHEP {\bf 1803}, 044 (2018)
  doi:10.1007/JHEP03(2018)044
  \href{https://arxiv.org/abs/1711.03901}{{\tt arXiv:1711.03901 [hep-th]}}.
  
\bibitem{ashoke} 
  A.~Laddha and A.~Sen,
  JHEP {\bf 1809}, 105 (2018)
  doi:10.1007/JHEP09(2018)105
  \href{https://arxiv.org/abs/1801.07719}{{\tt arXiv:1801.07719[hep-th]}}.
    
\bibitem{bern1} 
  Z.~Bern, T.~Dennen, Y.~t.~Huang and M.~Kiermaier,
  ``Gravity as the Square of Gauge Theory,''
  Phys.\ Rev.\ D {\bf 82}, 065003 (2010)
  doi:10.1103/PhysRevD.82.065003
  \href{https://arxiv.org/abs/1004.0693}{{\tt arXiv:1004.0693 [hep-th]}}.
  
\bibitem{bcf} 
  R.~Britto, F.~Cachazo and B.~Feng,
  ``New recursion relations for tree amplitudes of gluons,''
  Nucl.\ Phys.\ B {\bf 715}, 499 (2005)
  doi:10.1016/j.nuclphysb.2005.02.030
  \href{https://arxiv.org/abs/hep-th/0412308}{{\tt arXiv:hep-th/0412308}}.

 \bibitem{witten} 
  R.~Britto, F.~Cachazo, B.~Feng and E.~Witten,
  ``Direct proof of tree-level recursion relation in Yang-Mills theory,''
  Phys.\ Rev.\ Lett.\  {\bf 94}, 181602 (2005)
  doi:10.1103/PhysRevLett.94.181602
\href{https://arxiv.org/abs/hep-th/0501052}{{\tt arXiv:hep-th/0501052}}.
  
\bibitem{vera} 
  A.~Sabio Vera, E.~Serna Campillo and M.~A.~Vazquez-Mozo,
  ``Color-Kinematics Duality and the Regge Limit of Inelastic Amplitudes,''
  JHEP {\bf 1304}, 086 (2013)
  doi:10.1007/JHEP04(2013)086
  \href{https://arxiv.org/abs/1212.5103}{{\tt arXiv:1212.5103 [hep-th]}}.
  
\bibitem{johansson} 
  H.~Johansson, A.~Sabio Vera, E.~Serna Campillo and M.~Á.~Vázquez-Mozo,
  ``Color-Kinematics Duality in Multi-Regge Kinematics and Dimensional Reduction,''
  JHEP {\bf 1310}, 215 (2013)
  doi:10.1007/JHEP10(2013)215
  \href{https://arxiv.org/abs/1307.3106}{{\tt arXiv:1307.3106 [hep-th]}}.
  
\bibitem{johanssonochirov} 
H.~Johansson and A.~Ochirov,
``Pure Gravities via Color-Kinematics Duality for Fundamental Matter,''
JHEP {\bf 1511}, 046 (2015)
doi:10.1007/JHEP11(2015)046
\href{https://arxiv.org/abs/1407.4772}{{\tt arXiv:1407.4772 [hep-th]}}.
  
\bibitem{ochirov} 
  H.~Johansson and A.~Ochirov,
  ``Color-Kinematics Duality for QCD Amplitudes,''
  JHEP {\bf 1601}, 170 (2016)
  doi:10.1007/JHEP01(2016)170
  \href{https://arxiv.org/abs/1507.00332}{{\tt arXiv:1507.00332 [hep-th]}}.
  
\bibitem{ochirov2} 
  H.~Johansson and A.~Ochirov,
  ``Double copy for massive quantum particles with spin,''
\href{https://arxiv.org/abs/1906.12292}{{\tt arXiv:1906.12292 [hep-th]}}.
  
\bibitem{melia} 
T.~Melia,
``Dyck words and multiquark primitive amplitudes,''
Phys.\ Rev.\ D {\bf 88}, no. 1, 014020 (2013)
doi:10.1103/PhysRevD.88.014020
\href{https://arxiv.org/abs/1304.7809}{{\tt arXiv:1304.7809 [hep-ph]}}
  
\bibitem{goldberger1} 
  W.~D.~Goldberger, S.~G.~Prabhu and J.~O.~Thompson,
  ``Classical gluon and graviton radiation from the bi-adjoint scalar double copy,''
  Phys.\ Rev.\ D {\bf 96}, no. 6, 065009 (2017)
  doi:10.1103/PhysRevD.96.065009
  \href{https://arxiv.org/abs/1705.09263}{{\tt arXiv:1705.09263 [hep-th]}}.
  
\bibitem{li} 
  W.~D.~Goldberger, J.~Li and S.~G.~Prabhu,
  ``Spinning particles, axion radiation, and the classical double copy,''
  Phys.\ Rev.\ D {\bf 97}, no. 10, 105018 (2018)
  doi:10.1103/PhysRevD.97.105018
  \href{https://arxiv.org/abs/1712.09250}{{\tt arXiv:1712.09250 [hep-th]}}.
  
\bibitem{prabhu} 
  J.~Li and S.~G.~Prabhu,
  ``Gravitational radiation from the classical spinning double copy,''
  Phys.\ Rev.\ D {\bf 97}, no. 10, 105019 (2018)
  doi:10.1103/PhysRevD.97.105019
  \href{https://arxiv.org/abs/1803.02405}{{\tt arXiv:1803.02405 [hep-th]}}.
   
\bibitem{berends} 
  F.~A.~Berends and W.~T.~Giele,
  ``Multiple Soft Gluon Radiation in Parton Processes,''
  Nucl.\ Phys.\ B {\bf 313}, 595 (1989).
  doi:10.1016/0550-3213(89)90398-2
  
\bibitem{biswajit} 
  B.~Sahoo and A.~Sen,
  ``Classical and Quantum Results on Logarithmic Terms in the Soft Theorem in Four Dimensions,''
  JHEP {\bf 1902}, 086 (2019)
  doi:10.1007/JHEP02(2019)086
\href{https://arxiv.org/abs/1808.03288}{{\tt arXiv:1808.03288 [hep-th]}}.
  
\bibitem{elvang} 
  H.~Elvang, M.~Hadjiantonis, C.~R.~T.~Jones and S.~Paranjape,
  ``Soft Bootstrap and Supersymmetry,''
  JHEP {\bf 1901}, 195 (2019)
  doi:10.1007/JHEP01(2019)195
\href{https://arxiv.org/abs/1806.06079}{{\tt arXiv:1806.06079 [hep-th]}}.  

\bibitem{mcLerran} 
  L.~D.~McLerran and R.~Venugopalan,
  ``Computing quark and gluon distribution functions for very large nuclei,''
  Phys.\ Rev.\ D {\bf 49}, 2233 (1994)
  doi:10.1103/PhysRevD.49.2233
\href{https://arxiv.org/abs/hep-ph/9309289}{{\tt arXiv:hep-ph/9309289 [hep-ph]}}.  
    
\bibitem{venu} 
  S.~Jeon and R.~Venugopalan,
  ``Random walks of partons in SU(N(c)) and classical representations of color charges in QCD at small x,''
  Phys.\ Rev.\ D {\bf 70}, 105012 (2004)
  doi:10.1103/PhysRevD.70.105012
\href{https://arxiv.org/abs/hep-ph/0406169}{{\tt arXiv:hep-ph/0406169 [hep-ph]}}.  
  
\bibitem{song} 
  S.~He, Y.~t.~Huang and C.~Wen,
  ``Loop Corrections to Soft Theorems in Gauge Theories and Gravity,''
  JHEP {\bf 1412}, 115 (2014)
  doi:10.1007/JHEP12(2014)115
\href{https://arxiv.org/abs/1405.1410}{{\tt arXiv:1405.1410 [hep-th]}}.  
  
\bibitem{white} 
  C.~D.~White,
  ``Diagrammatic insights into next-to-soft corrections,''
  Phys.\ Lett.\ B {\bf 737}, 216 (2014)
  doi:10.1016/j.physletb.2014.08.041
\href{https://arxiv.org/abs/1406.7184}{{\tt arXiv:1406.7184 [hep-th]}}  
  
\bibitem{vera1} 
  A.~Sabio Vera and M.~A.~Vazquez-Mozo,
  ``The Double Copy Structure of Soft Gravitons,''
  JHEP {\bf 1503}, 070 (2015)
  doi:10.1007/JHEP03(2015)070
  \href{https://arxiv.org/abs/1412.3699}{{\tt arXiv:1412.3699 [hep-th]}}.
    
\bibitem{oxburgh} 
  S.~Oxburgh and C.~D.~White,
  ``BCJ duality and the double copy in the soft limit,''
  JHEP {\bf 1302}, 127 (2013)
  doi:10.1007/JHEP02(2013)127
\href{https://arxiv.org/abs/1210.1110}{{\tt arXiv:1210.1110 [hep-th]}}.  
  
\bibitem{bernloop} 
  Z.~Bern, S.~Davies, T.~Dennen, Y.~t.~Huang and J.~Nohle,
  ``Color-Kinematics Duality for Pure Yang-Mills and Gravity at One and Two Loops,''
  Phys.\ Rev.\ D {\bf 92}, no. 4, 045041 (2015)
  doi:10.1103/PhysRevD.92.045041
  \href{https://arxiv.org/abs/1303.6605}{{\tt arXiv:1303.6605 [hep-th]}}.
  
\bibitem{bern2} 
  Z.~Bern, S.~Davies, P.~Di Vecchia and J.~Nohle,
  ``Low-Energy Behavior of Gluons and Gravitons from Gauge Invariance,''
  Phys.\ Rev.\ D {\bf 90}, no. 8, 084035 (2014)
  doi:10.1103/PhysRevD.90.084035
  \href{https://arxiv.org/abs/1406.6987}{{\tt arXiv:1406.6987 [hep-th]}}.
  
\bibitem{broedel} 
  J.~Broedel, M.~de Leeuw, J.~Plefka and M.~Rosso,
  ``Constraining subleading soft gluon and graviton theorems,''
  Phys.\ Rev.\ D {\bf 90}, no. 6, 065024 (2014)
  doi:10.1103/PhysRevD.90.065024
  \href{https://arxiv.org/abs/1406.6574}{{\tt arXiv:1406.6574 [hep-th]}}.
  
\bibitem{kovchegov} 
  Y.~V.~Kovchegov and D.~H.~Rischke,
  Phys.\ Rev.\ C {\bf 56}, 1084 (1997)
  doi:10.1103/PhysRevC.56.1084
\href{https://arxiv.org/abs/hep-ph/9704201}{{\tt arXiv:hep-ph/9704201 [hep-ph]}}.  
  
  
\end{thebibliography}
\end{document}